\documentstyle[12pt]{article}
\begin{document}
\addtolength{\topmargin}{-30pt}
\addtolength{\textheight}{70pt}
\oddsidemargin8mm
\newcommand{\gai}{\gamma_i}
\newcommand{\gaj}{\gamma_j}
\newcommand{\gak}{\gamma_k}
\newcommand{\gam}{\gamma_m}
\newcommand{\gaa}{\gamma_a}
\newcommand{\gab}{\gamma_b}
\newcommand{\gac}{\gamma_c}
\newcommand{\reals}{\mbox{${\rm I\!R }$}}
\newcommand{\nats}{\mbox{${\rm I\!N }$}}
\newcommand{\intgs}{\mbox{${\rm Z\!\!Z }$}}
\newcommand{\ptr}{{\rm Tr}}
\def\preal{\hbox{ }{{\vrule height.65em width.040em
     depth-.04em}\kern-0.35em{\rm R}}}
\def\birdy{\textstyle{d\over{d\epsilon}}|_{\epsilon=0}}

\def\qedbox{\hbox{$\rlap{$\sqcap$}\sqcup$}}
\def\qed{\nobreak\hfill\penalty250 \hbox{}\nobreak\hfill\qedbox}
\def\proof{\medbreak\noindent{\bf Proof:} \rm }
\def\beq{\begin{eqnarray}}
\def\eeq{\end{eqnarray}}
\newcommand{\nn}{\nonumber}
\newcommand{\PGa}{\qquad}
\newcommand{\PGc}{\qquad\qquad}
\newcommand{\PGb}{\qquad\quad}
\newcommand{\sujeu}{\sum_{j=1}^{\infty}}
\newcommand{\sujnu}{\sum_{j=0}^{\infty}}
\newcommand{\sulnu}{\sum_{l=0}^{\infty}}
\newcommand{\suani}{\sum_{a=0}^i}
\newcommand{\sneu}{\sum_{n=1}^\infty}
\newcommand{\snnu}{\sum_{n=0}^\infty}
\newcommand{\slnu}{\sum_{l=0}^\infty}
\newcommand{\sluu}{\sum_{l=-\infty}^\infty}
\newcommand{\svecn}{\sum_{\vec n \in \intgs^d}}
\newcommand{\svecno}{\sum_{\vec n \in \intgs^d /\{\vec 0\}}}
\newcommand{\res}{{\rm Res}\phantom{,}}
\newcommand{\Res}{{\rm Res\,\,}}
\newcommand{\pa}{\partial}
\newcommand{\zend}{\zeta^{\nu}}
\newcommand{\zb}{\zeta_{{\cal N}}}
\newcommand{\zn}{\zeta_{{\cal N}} \left(}
\newcommand{\zh}{\zeta_H}
\newcommand{\zr}{\zeta_R}
\newcommand{\zba}{\zeta_{{\cal B}}}
\newcommand{\zeb}{\zeta_{{\cal B}}}
\newcommand{\zem}{\zeta_{{\cal M}}}
\newcommand{\pold}{D^{(d-1)}}
\newcommand{\hem}{A^{\cam}}
\newcommand{\hen}{A^{\can}}
\newcommand{\facb}{\frac{(4\pi)^{d/2}}{a^d|S^d|}}
\newcommand{\fac}{\frac{(4\pi)^{D/2}}{a^d|S^d|}}
\newcommand{\sip}{\frac{\sin (\pi s)}{\pi}}
\newcommand{\mzs}{m^{-2s}}
\newcommand{\g}{\Gamma\left(}
\newcommand{\numr}{\left(\frac{\nu}{mR}\right)^2}
\newcommand{\rzs}{R^{2s}}
\newcommand{\lll}{\frac{(-1)^j}{j!}}
\newcommand{\ent}{d (\nu )}
\def\ball{{\rm ball}}
\newcommand{\ph}{\phantom}
\newcommand{\cab}{{\cal B}} 
\newcommand{\cac}{{\cal C}}
\newcommand{\can}{{\cal N}}
\newcommand{\cam}{{\cal M}}
\newcommand{\man}{{\cal M}}
\newtheorem{theorem}{Theorem}[section]
\newtheorem{corollary}[theorem]{Corollary}
\newtheorem{lemma}[theorem]{Lemma}
\newtheorem{proposition}[theorem]{Proposition}
\newtheorem{definition}[theorem]{Definition}
\newtheorem{example}[theorem]{Example}
\newcommand{\al}{\alpha}
\newcommand{\be}{\beta}
\newcommand{\de}{\delta}
\newcommand{\ep}{\epsilon}
\newcommand{\ga}{\gamma}
\newcommand{\la}{\lambda}
\newcommand{\om}{\omega}
\newcommand{\ze}{\zeta}
\def\text#1{{\hbox{#1}}}
\newcommand{\De}{\Delta}
\newcommand{\Om}{\Omega}
\newcommand{\Si}{\Sigma}
\newcommand{\G}{\Gamma}
\newcommand{\Ga}{\Gamma}
\newcommand{\La}{\Lambda}
\newcommand{\Th}{\Theta}
\def\pha{\phantom a}
\def\ca{\chi_{:a}}
\def\cb{\chi_{:b}}
\def\fm{F_{;m}}
\def\fmm{F_{;mm}}
\def\am{f_{,m}}
\def\amm{f_{,mm}}
\def\aam{f_{,am}}
\def\et{\medbreak\noindent\rm}
\def\fract/#1/#2/{{\textstyle{{#1}\over{#2}}}}
\newcommand{\noin}{\noindent}
\def\pbg{\par\vglue 0cm\ \phantom{aaaa}}
\setcounter{page}{1}
\setcounter{section}{0}
\def\nmonth{\ifcase\month\ \or January\or
   February\or March\or April\or May\or June\or July\or August\or
   September\or October\or November\else December\fi}
\def\version{5e}
\title{Heat kernel asymptotics with mixed boundary conditions
}
\author{Thomas P.~Branson, Peter B.~Gilkey\footnote
   {Research partially supported by the ESI (Austria), IHES (France), 
    and the NSF (USA)},\\
 Klaus Kirsten\footnote{Research partially supported by the DFG under 
contract number Bo1112/4-2 and the EPSRC under Grant No GR/M45726}, and
Dmitri V.~Vassilevich\footnote{Research partially supported
by the Alexander von Humboldt foundation (Germany), ICTP (Italy)
and GRACENAS (Russia)}
}
\maketitle
\begin{abstract}
We calculate the coefficient $a_5$ of the heat kernel asymptotics
for an operator of Laplace type with mixed boundary conditions
on a general compact manifold.
\end{abstract}

\section{Introduction}\label{S1}
Let $V$ be a vector bundle over a smooth
compact Riemannian manifold $M$ of dimension $m$ with smooth boundary
$\partial M$. Let $D$ be a second order operator of Laplace type on $C^\infty(V)$;
this means the leading symbol of $D$ is given by the metric tensor, or equivalently,
that $D$ has the form given in equations (\ref{laplacea}) and (\ref{laplaceb}) below. Note that many
natural second order operators which arise in applications are of Laplace type.
If the boundary is non-empty, then we must impose suitable boundary conditions. We assume given a
decomposition
$V|_{\partial M}=V_N\oplus V_D$; extend the decomposition
to be parallel with respect to the normal geodesic rays on a collared
neighborhood of the boundary. Let
$S$ be an auxiliary endomorphism of the bundle
$V_N$ and let $\phi_{;m}$ be the covariant
derivative of $\phi\in C^\infty(V)$ with respect to the inward
unit normal.  Decompose $\phi=\phi^N\oplus\phi^D$ and set
$$\cab\phi:=\phi^D|_{V_D}\oplus(\phi^N_{;m}+S\phi^N)|_{V_N}.$$
Dirichlet boundary conditions correspond to vanishing $V_N$; Robin (i.e.\
modified Neumann) boundary conditions correspond to vanishing $V_D$.  Let
$D_\cab$ be the realization of
$D$; $\phi\in\text{Domain}(D_\cab)$ if and only if
$\cab\phi=0$. This is an elliptic
boundary value problem and satisfies the Lopatinski-Shapiro condition.

Let $F$ be an auxiliary smooth function on $M$, for use in localization or
`smearing'. Let
$e^{-tD_\cab}$ be the fundamental solution of the heat equation; this is
trace class on $L^2(V)$ and as $t\downarrow 0$, there is an asymptotic
expansion:
$$\ptr_{L^2}(Fe^{-tD_\cab})\cong\sum_{n\ge0}t^{(n-m)/2}a_n(F,D,\cab);$$
see \cite{greiner, grubb, seeley} for
details. The invariants  $a_n(F,D, \cab )$ are locally computable in terms of
geometric invariants and form the focus of our study.

The study of these invariants can be physically motivated. Consider a
field theory in which $D$ defines the quadratic form of the action; see, for example \cite{EKP}. We note
that consistent local boundary conditions for fields of non-zero spin are inevitably mixed ones.
The one-loop effective action is (formally) given by $\log \det (D)$. The heat kernel
coefficients up to $a_m$ describe ultraviolet divergencies of the theory. The coefficient
$a_m$ also defines the scale anomaly. The leading term in the large mass expansion of the effective
action is given by
$a_{m+1}$; the lower coefficients are usually absorbed in renormalizations.
Variations of the heat kernel coefficients give vacuum expectation values of
currents, see \cite{BGV98} for example. The study of these coefficients is
important for renormalization and calculation of anomalies in higher
dimensional models
\cite{S98}, and for large mass expansion and calculation of currents
in four dimensions.

McKean and Singer \cite{MS} studied the invariants $a_0$, $a_1$, and $a_2$.
Kennedy et al.\ \cite{KCD} studied the invariant $a_3$. Moss and Dowker
\cite{MD} studied
$a_4$ for the scalar Laplacian. Branson and Gilkey \cite{BG90} determined $a_4$
in the vector valued case; a minor error in the calculation of two of the
coefficients was later corrected by Vassilevich \cite{V95}. These results are
summarized in Theorem \ref{ThmA}. Many other authors have also studied
these coefficients. 

We say the boundary conditions are {\em pure} if $V_N$ or $V_D$ vanishes. The
coefficient $a_5$ has been determined previously for pure boundary conditions. Branson,
Gilkey and Vassilevich
\cite{BGV95} studied the special cases of a domain $M$ in flat space
and of a curved domain with totally geodesic boundary. Kirsten \cite{K98}
generalized these results to arbitrary manifolds and boundaries; see also
Dowker and Kirsten
\cite{smeared} for related work. We summarize the results of those papers in
Lemmas \ref{EQGb} and \ref{EQGc}.  The main result of this paper is Theorem 
\ref{ThmB} which gives $a_5$ for general mixed boundary
conditions.

We have decoupled the calculation of $a_5$ in the general
setting into three pieces. The first piece,
${\cal A}_5^1$, was computed in \cite{BGV95}; it gives  $a_5$ for pure
boundary conditions if the boundary is totally geodesic. The second piece
${\cal A}_5^2$ was computed in \cite{K98}; it contains the terms
involving the second fundamental form which are necessary to deal with the case
where  the boundary is not totally geodesic \cite{K98}. The final piece
${\cal A}_5^3$ is computed in this paper; it contains the additional 42 terms which describe
the interaction of $V_N$ and $V_D$ for general mixed boundary conditions. Our
purpose in this paper is not purely combinatorial; determining these additional coefficients
requires us to derive new functorial properties of the invariants which are
important in their own right. 

In \S\ref{Sectc}, we use invariance theory to show that the computation of
$a_5$ can be decomposed into the three pieces discussed above and to show
that the undetermined coefficients $w_i$ describing the interaction of
$V_N$ and $V_D$ are universal coefficients which are dimension free.
The remainder of the paper is devoted to computing these coefficients; see
the cross reference table following Theorem \ref{ThmB}. We postpone until
Appendices A, B, and C the discussion of some combinatorial formulas we
shall need.

We use three different technical tools in the proof of Theorem \ref{ThmB}. 
\begin{enumerate}
\item \bf The index
theorem. \rm Let $P:C^\infty(V)\rightarrow C^\infty(W)$ be an elliptic
complex of Dirac type. Impose suitable
boundary conditions. Let
$D:=P^*P$ and
$\hat D:=PP^*$ be the associated operators of Laplace type. An observation
of Bott shows that with suitable boundary conditions
$a_m(1,D,\cab)-a_m(1,\hat D,\cab)=\text{index}(P)$ and $a_n(1,D,\cab)-a_n(1,\hat D,\cab)=0$
if $n\ne m$; this is often called the local index formula. We shall apply this
observation in \S\ref{Sectd} and in
\S\ref{Sectg}.

\item \bf Conformal variations. \rm Let $ D( \epsilon )=e^{ -2 \epsilon
F}D$. We then have that $  \birdy a_n(1,D, \cab )=(m-n)a_n(F,D, \cab )$.
Furthermore, if
$ m=n+2$, then
$  \birdy a_n(e^{ -2 \epsilon f}F,e^{-2 \epsilon f}D, \cab )=0$.
We use these identities in \S\ref{Sectf}.

\item \bf Calculations on the ball. \rm The invariant $a_n$ was computed
for the Laplacians on spinors and on $1$-forms on the ball in 
\cite{DAKB96,stukla}; we use this computation in
\S\ref{Secte}.\end{enumerate}

\section{Statement of results}\label{Sectb}

If $D$ is an operator of Laplace type, then locally we have:
\begin{equation}D=-(g^{\mu\nu}\partial_\mu\partial_\nu+A^\sigma\partial_\sigma+B).
\label{laplacea}\end{equation}
Here $g^{\mu\nu}$ acts by scalar multiplication; $A^\sigma$ and $B$ are matrix valued. We can also
express
$D$ covariantly. There is a unique connection
$\nabla$ on
$V$ and a unique endomorphism $E$ of $V$ so that
\begin{equation}D=-(\ptr(\nabla^2)+E).\label{laplaceb}\end{equation}
Let $\omega$ be
the
local $1$-form of the connection
$\nabla$. We may express:
\begin{eqnarray*}
    &&\omega_\delta=\fract/1/2/g_{\nu\delta}(A^\nu+g^{\mu\sigma}\Gamma_{\mu
    \sigma}{}^{\nu }I_{V})\text{ and}\\
    &&E=B-g^{\nu \mu}(
    \partial_\mu\omega_\nu+\omega_\nu\omega_\mu
     -\omega_\sigma\Gamma_{\nu\mu}{}^\sigma).\end{eqnarray*}
If $D$ is the
scalar Laplacian, then $\nabla$ is the flat connection and $E=0$. If $D=d\delta+\delta d$ on the
bundle of
$p$ forms, then $\nabla$ is Levi-Civita connection and the endomorphism $E$ is given in
terms of curvature by the Weitzenb\"ock formulas. Let $\tau$ be the scalar
curvature. If $D$ is the spin Laplacian, then $\nabla$ is the spin connection,
and
the Lichnerowicz formula gives $E=-{1\over4}\tau$.

We adopt the following notational conventions to express the asymptotic
coefficients $a_n$ in terms of geometrical invariants. Let
Roman indices $i$, $j$, $k$, and $l$ range from 1 through
the dimension $m$ of the manifold and index a local orthonormal frame
$\{ e_1,...,e_m\}$ for the tangent bundle of the manifold. Let Roman indices
$a$, $b$, $c$, and $d$ range from 1 through $m-1$ and index a local orthonormal
frame for the tangent bundle of the boundary
$\partial M$. On the boundary, we shall let $e_m$ be the inward unit normal
vector field. We shall not introduce bundle indices explicitly. Let $\ptr({\cal A})$
be the fiber trace of an endomorphism $\cal A$ of $V$. Greek indices will
index coordinate frames. We adopt the Einstein convention and sum over repeated indices.

Let $R_{ijkl}$ be the components of the curvature tensor of the
Levi-Civita connection, let $\rho_{ij}:=R_{ikkj}$ be the components of the
Ricci tensor, and let $\tau:=\rho_{ii}$ be the scalar curvature. With our sign
convention, $R_{1212}=-1$ on the unit sphere $S^2$ in Euclidean space. Let
$\Omega_{ij}$ be the endomorphism-valued components of the curvature of the connection
$\nabla$ on $V$. Let $\Gamma$ be the Christoffel
symbols of the Levi-Civita connection. Let $L_{ab}:=(\nabla_{e_{a}}e_{b},e_{m}) =
\Gamma_{abm}$ give the second fundamental form. We use the Levi-Civita connections and
the connection $\nabla$ to covariantly differentiate tensors of all types. Let
\lq ;\rq\ denote multiple covariant differentiation with respect to the Levi-Civita
connection of
$M$ and let
\lq:\rq\ denote multiple tangential covariant differentiation on the boundary with respect to the
Levi-Civita connection of the boundary; the difference between `;' and `:' is measured by the second
fundamental form. Thus, for example, $E_{;a}=E_{:a}$ since there are no tangential indices in $E$ to be
differentiated. On the other hand, 
$E_{;ab}\ne E_{:ab}$ since the index $a$ is also being
differentiated. Since $L$ and $S$ are only defined on the boundary, these
tensors can only be differentiated tangentially.
Let $dx$ and $dy$ be the Riemannian volume elements on
$M$ and on $\partial M$ respectively. Let $f_1\in C^{\infty}(M)$ and let
$f_{2}\in C^{\infty }(\partial M)$. Let
$$
   f_1[M]:={\textstyle\int}_Mf_1(x)dx\text{ and }
   f_{2}[\partial M]={\textstyle\int}_{\partial M}f_{2}(y)dy.
$$

Let $\Pi_-$ be orthogonal projection on $V_D$ and let $\Pi_+$ be
orthogonal projection on $V_N$. Let $\chi:=\Pi_+-\Pi_-$; $\chi=+1$ on $V_N$ and $\chi=-1$ on
$V_D$. Extend $\chi$ to be parallel along normal geodesic rays. We refer to \cite{BG90} for the
proof of the following result:

\begin{theorem}\label{ThmA}\ \begin{enumerate}
\item $a_0(F,D,\cab)=(4\pi)^{-m/2}\ptr(F)[M]$.
\item $a_1(F,D,\cab)={1\over4}(4\pi)^{-(m-1)/2}
     \ptr(\chi F)[\partial M]$.
\item $a_2(F,D,\cab)=\fract/1/6/(4\pi)^{-m/2}\{
     \ptr(6FE+F\tau)[M]$\pbg
    $+\ptr(2FL_{ aa}+3 \chi F_{ ;m}+12FS)[\partial M]\}$.
\item $a_3(F,D,\cab)=\fract/1/384/(4 \pi )^{
       -(m-1)/2}  \ptr\big\{ F(96 \chi E+16\chi  \tau +8F \chi
       R_{amam}$\pbg
        $+(13 \Pi_{+}-7 \Pi_{ -})L_{ aa}L_{ bb}+(2 \Pi_{ +}+10
      \Pi_{-})L_{ab}L_{ ab}+96SL_{ aa}
      +192S^2$\pbg$-12 \chi_{ :a} \chi_{:a})+F_{ ;m}((6 \Pi_{ +}+30 \Pi_{ -})L_{
        aa}+96S)+24 \chi F_{ ;mm} \} [\partial M]$.
\item $a_4(F,D,\cab)=\fract/1/360/(4 \pi )^{
       -m/2} \big\{\ptr
       \{ F(60E_{ ; ii}+60 \tau E+180E^2$\pbg
       $+30 \Omega_{ij} 
      \Omega_{ij} +12 \tau_{;ii}+5 \tau^2-2\rho_{ij}\rho_{ij}+2R_{ijkl}R_{ijkl}) \}
[M]$\pbg
     $+ \ptr \big\{ F \{ (240 \Pi_{ +}-120 \Pi_{ -})E_{ ;m}+(42 \Pi_{
     +}-18 \Pi_{ -}) \tau_{ ;m} $\pbg$+24L_{ aa:bb}+0L_{ ab:ab}+120EL_{
       aa}+20 \tau L_{ aa}+4R_{ am    am}L_{
bb}$\pbg$-12R_{ am    bm}L_{ ab}+4R_{
ab    cb}L_{ ac}+0 \Omega_{
am:a}+\fract/1/21/ \{ (280 \Pi_{ +}+40 \Pi_{
-})L_{ aa}L_{ bb}L_{ cc}$\pbg$
    +(168 \Pi_{ +}-264 \Pi_{
-})L_{ ab}L_{ ab}L_{ cc}+(224 \Pi_{
+}+320 \Pi_{ -})L_{ ab}L_{ bc}L_{
ac} \}$\pbg$+720SE+120S \tau +0SR_{ am    am}+144SL_{
aa}L_{ bb}+48SL_{ ab}L_{ ab}$\pbg$+480S^2L_{ aa}+480S^{
3}+120S_{ :aa}+60 \chi  \chi_{ :a} \Omega_{ am}-12 \chi_{ :a} \chi_{
:a}L_{ bb}$\pbg$-24 \chi_{ :a} \chi_{
:b}L_{ ab}-120 \chi_{ :a} \chi_{
:a}S \} +F_{ ;m}(180 \chi E+30 \chi  \tau +0R_{
am    am}$\pbg$+\fract/1/7/ \{ (84 \Pi_{
+}-180 \Pi_{ -})L_{ aa}L_{ bb}+(84 \Pi_{
+}+60 \Pi_{ -})L_{ ab}L_{ ab} \} $\pbg$+72SL_{ aa}+240S^2-18 \chi_{
:a} \chi_{ :a})+F_{ ;mm}(24L_{
aa}+120S)$\pbg$+30 \chi F_{ ;iim}  \big\} [\partial M]\}$.
\end{enumerate}
\end{theorem}

We introduce some additional notation to discuss $a_5$. Let:
\medbreak\qquad 
    ${\cal A}_5^1:=F\{360\chi E_{;mm} +1440 E_{;m} S$
\smallbreak\qquad\qquad\quad 
     $+720\chi E^2+240\chi  E_{:aa}
      +240\chi \tau E+48\chi\tau_{;ii}
      +20\chi\tau^2$
\smallbreak\qquad\qquad\quad
     $-8\chi\rho_{ij}\rho_{ij}+8\chi R_{ijkl}R_{ijkl} 
      -120\chi\rho_{mm}E-20\chi\rho_{mm}\tau$
\smallbreak\qquad\qquad\quad 
     $+480 \tau S^2+12\chi \tau_{;mm}
      +24\chi \rho_{mm:aa}+15\chi\rho_{mm;mm}$
\smallbreak\qquad\qquad\quad
      $+270\tau_{;m}S+120 \rho_{mm}S^2
       +960SS_{:aa}+16\chi R_{ammb}\rho_{ab}$
\smallbreak\qquad\qquad\quad 
      $-17\chi \rho_{mm}\rho_{mm}-10\chi R_{ammb}R_{ammb}
       +2880ES^2+1440S^4\}$
\smallbreak\qquad\qquad
      $+F_{;m}\{(\frac{195}2\Pi_+-60\Pi_-)\tau_{;m} 
       +240\tau S-90\rho_{mm}S+270 S_{:aa}$
\smallbreak\qquad\qquad\quad
      $+(630\Pi_+-450\Pi_-)E_{;m}+1440 ES+720S^3\}$
\smallbreak\qquad\qquad
      $+F_{;mm}\{60\chi\tau -90\chi\rho_{mm} 
       +360\chi E +360 S^2\}$
\smallbreak\qquad\qquad
       $+180SF_{;mmm}+45\chi F_{;mmmm}$.
\medbreak\noindent
Let $\cab_S^-$ denote pure
Dirichlet boundary conditions if $V_N$ vanishes; $S$ plays no role here. Similarly let
$\cab_S^+$ denote Neumann boundary conditions modified by
$S$ if $V_D$ vanishes. The following result was
proved in
\cite{BGV95}.

\begin{lemma}\label{EQGb} If the boundary of $M$ is totally
geodesic, then we have
\begin{eqnarray*}
     &&a_5(F,D,\cab_S^\pm)=\fract/1/5760/(4\pi)^{-(m-1)/2}\ptr\{{\cal A}_5^1
    +120F\chi\Omega_{ab}\Omega_{ab}\\
     &&\qquad
      +(90\Pi_++360\Pi_-)F\Omega_{am}
     \Omega_{am}+600FS_{:a}S_{:a}\}[\partial M].\nonumber\end{eqnarray*}\end{lemma}

Next we introduce terms involving the second fundamental form $L$. Let:
\medbreak\qquad ${\cal A}_5^2:=F\{(90\Pi_++450\Pi_-)L_{aa}E_{;m} 
  +(\fract/111/2/\Pi_++42\Pi_-)L_{aa}\tau_{;m}$
\smallbreak\qquad\qquad\quad
  $+30\Pi_+L_{ab}R_{ammb;m}
  +240L_{aa}S_{:bb}
  +420L_{ab}S_{:ab} 
  +390L_{aa:b}S_{:b}$
\smallbreak\qquad\qquad\quad
  $+480L_{ab:a}S_{:b}
  +420L_{aa:bb}S +60L_{ab:ab}S$
\smallbreak\qquad\qquad\quad
  $+(\fract/487/16/\Pi_+ +\fract/413/16/\Pi_- ) L_{aa:b} L_{cc:b}
  +(238\Pi_+-58\Pi_-)L_{ab:a}L_{cc:b}$
\smallbreak\qquad\qquad\quad
  $+(\fract/49/4/\Pi_++\fract/11/4/\Pi_-)L_{ab:a}L_{bc:c}
   +(\fract/535/8/\Pi_+-\fract/355/8/\Pi_-)L_{ab:c}L_{ab:c}$
\smallbreak\qquad\qquad\quad
  $+(\fract/151/4/\Pi_++\fract/29/4/\Pi_-) L_{ab:c} L_{ac:b}
  +(111\Pi_+-6\Pi_-)L_{aa:bb}L_{cc}$
\smallbreak\qquad\qquad\quad
  $+(-15\Pi_++30\Pi_-) L_{ab:ab}L_{cc}
  +(-\fract/15/2/\Pi_++\fract/75/2/\Pi_-)L_{ab:ac}L_{bc}$
\smallbreak\qquad\qquad\quad
  $+(\fract/945/4/\Pi_+-\fract/285/4/\Pi_-)L_{aa:bc}L_{bc} 
  +(114\Pi_+-54\Pi_-) L_{bc:aa}L_{bc}$
\smallbreak\qquad\qquad\quad
  $+1440L_{aa}SE+30L_{aa}S\rho_{mm}+240L_{aa}S\tau
  -60L_{ab}\rho_{ab}S$
\smallbreak\qquad\qquad\quad
  $+180L_{ab}SR_{ammb}
  +(195\Pi_+-105\Pi_-)L_{aa}L_{bb}E$
\smallbreak\qquad\qquad\quad
  $+(30\Pi_+ +150\Pi_-) L_{ab}L_{ab}E
  +(\fract/195/6/\Pi_+-\fract/105/6/\Pi_-)L_{aa}L_{bb}\tau$
\smallbreak\qquad\qquad\quad
  $+(5\Pi_++25\Pi_-)L_{ab}L_{ab}\tau
  +(-\fract/275/16/\Pi_++\fract/215/16/\Pi_-)L_{aa}L_{bb}\rho_{mm}$
\smallbreak\qquad\qquad\quad
  $+(-\fract/275/8/\Pi_++\fract/215/8/\Pi_-)L_{ab}L_{ab}\rho_{mm}
  +(-\Pi_+-14\Pi_-)L_{cc}L_{ab}\rho_{ab}$
\smallbreak\qquad\qquad\quad
  $+(\fract/109/4/\Pi_+-\fract/49/4/\Pi_-)L_{cc}L_{ab}R_{ammb}
  +16\chi L_{ab}L_{ac}\rho_{bc}$
\smallbreak\qquad\qquad\quad
  $+(\fract/133/2/\Pi_++\fract/47/2/\Pi_-)L_{ab}L_{ac}R_{bmmc}
  -32\chi L_{ab}L_{cd} R_{acbd}$
\smallbreak\qquad\qquad\quad
  $+\fract/315/2/L_{cc}L_{ab}L_{ab}S
  +(\fract/2041/128/\Pi_+ +\fract/65/128/\Pi_-)L_{aa}L_{bb}L_{cc}L_{dd}$
\smallbreak\qquad\qquad\quad
  $+150L_{ab}L_{bc}L_{ac}S
  +(\fract/417/32/\Pi_++\fract/141/32/\Pi_-)L_{cc}L_{dd}L_{ab}L_{ab}$
\smallbreak\qquad\qquad\quad 
  $+1080L_{aa}L_{bb}S^2
  +360L_{ab}L_{ab}S^2
  +(\fract/375/32/\Pi_+ -\fract/777/32/\Pi_-)L_{ab}L_{ab}L_{cd}L_{cd}$
\smallbreak\qquad\qquad\quad
  $+\fract/885/4/L_{aa}L_{bb}L_{cc}S
   +(25\Pi_+-\fract/17/2/\Pi_-)L_{dd}L_{ab}L_{bc}L_{ac}$
\smallbreak\qquad\qquad\quad
  $+2160L_{aa}S^3
  +(\fract/231/8/\Pi_+ +\fract/327/8/\Pi_-) L_{ab}L_{bc}L_{cd}L_{da}\}$
\smallbreak\qquad\qquad
  $+F_{;m}\{(90\Pi_++450\Pi_-)L_{aa}E
  +(-\fract/165/8/\Pi_+-\fract/255/8/\Pi_-)L_{aa}\rho_{mm}$
\smallbreak\qquad\qquad\quad
  $+(15\Pi_++75\Pi_-)L_{aa}\tau
  +600L_{aa}S^2+(\fract/1215/8/\Pi_+-\fract/315/8/\Pi_-)L_{aa:bb}$
\smallbreak\qquad\qquad\quad
  $-\fract/45/4/\chi L_{ab:ab}
  +(15\Pi_+-30\Pi_-)L_{ab}\rho_{ab}$
\smallbreak\qquad\qquad\quad
  $+(-\fract/165/4/\Pi_++\fract/465/4/\Pi_-)L_{ab}R_{ammb}
  +\fract/705/4/L_{aa}L_{bb}S
  -\fract/75/2/L_{ab}L_{ab}S$
\smallbreak\qquad\qquad\quad
  $+(\fract/459/32/\Pi_++\fract/495/32/\Pi_-)L_{aa}L_{bb}L_{cc}
   +(\fract/267/16/\Pi_+-\fract/1485/16/\Pi_-)L_{cc}L_{ab}L_{ab}$
\smallbreak\qquad\qquad\quad
  $+(-54\Pi_++\fract/225/2/\Pi_-)L_{ab}L_{bc}L_{ac}\}$
\smallbreak\qquad\qquad
  $+F_{;mm}\{30L_{aa}S+(\fract/315/16/\Pi_+-\fract/1215/16/\Pi_-)L_{aa}L_{bb}$
\smallbreak\qquad\qquad\quad
  $+(-\fract/645/8/\Pi_++\fract/945/8/\Pi_-)L_{ab}L_{ab}$
\smallbreak\qquad\qquad
  $+F_{;mmm}(-30\Pi_++105\Pi_-)L_{aa}$.
\medbreak\noindent The following result was proved in \cite{K98}.
\begin{lemma}\label{EQGc} We have:
\begin{eqnarray*}
     &&a_5(F,D,\cab_S^\pm)=\fract/1/5760/(4\pi)^{-(m-1)/2}\ptr\{{\cal A}_5^1
    +{\cal A}_5^2
    +120F\chi\Omega_{ab}\Omega_{ab}\\
     &&\qquad
      +(90\Pi_++360\Pi_-)F\Omega_{am}
     \Omega_{am}+600FS_{:a}S_{:a}\}[\partial M].\nonumber\end{eqnarray*}\end{lemma}

To generalize Lemma \ref{EQGc} to mixed boundary conditions, we
introduce the additional terms describing the interaction of $V_N$ and $V_D$. Let:
\medbreak\qquad ${\cal A}_5^3(\vec w):=F\{w_1E^2+w_2\chi E\chi E+w_3S_{:a}S_{:a}
    +w_4\chi S_{:a}S_{:a}$
\smallbreak\qquad\qquad\quad
    $+w_5\Omega_{ab}\Omega_{ab}
    +w_6 \chi\Omega_{ab}\Omega_{ab} + w_7\chi\Omega_{ab}\chi\Omega_{ab} 
    +w_8\Omega_{am}\Omega_{am}$
\smallbreak\qquad\qquad\quad
    $+w_9\chi\Omega_{am}\Omega_{am}
    +w_{10}\chi\Omega_{am}\chi\Omega_{am}
    +w_{11}(\Omega_{am}\chi S_{:a}-\Omega_{am}S_{:a}\chi)$
\smallbreak\qquad\qquad\quad
    $+w_{12}\chi\chi_{:a}\Omega_{am}L_{cc}+w_{13}\chi_{:a}\chi_{:b}\Omega_{ab}
    +w_{14}\chi\chi_{:a}\chi_{:b}\Omega_{ab}$
\smallbreak\qquad\qquad\quad
    $+w_{15}\chi\chi_{:a}\Omega_{am;m}
    +w_{16}\chi\chi_{:a}\Omega_{ab:b}
    +w_{17}\chi\chi_{:a}\Omega_{bm}L_{ab} 
    +w_{18}\chi_{:a} E_{:a}$
\smallbreak\qquad\qquad\quad 
    $+w_{19}\chi_{:a}\chi_{:a}E+w_{20}\chi\chi_{:a}\chi_{:a}E+w_{21}\chi_{:aa}E
    +w_{22}\chi_{:a}\chi_{:a}\tau$
\smallbreak\qquad\qquad\quad
    $+w_{23}\chi_{:a}\chi_{:a}\rho_{mm}
    +w_{24}\chi_{:a}\chi_{:b}\rho_{ab}+w_{25}\chi_{:a}\chi_{:b}R_{mabm}$
\smallbreak\qquad\qquad\quad
   $+w_{26}\chi_{:a}\chi_{:a}L_{bb}L_{cc}
    +w_{27}\chi_{:a}\chi_{:b}L_{ac}L_{bc}
    +w_{28}\chi_{:a}\chi_{:a}L_{cd}L_{cd}$
\smallbreak\qquad\qquad\quad
    $+w_{29}\chi_{:a}\chi_{:b}L_{ab}L_{cc}
    +w_{30}\chi_{:a}S_{:a}L_{cc}+w_{31}\chi_{:a}S_{:b}L_{ab}$
\smallbreak\qquad\qquad\quad
    $+w_{32}\chi_{:a}\chi_{:a}\chi_{:b}\chi_{:b}
    +w_{33}\chi_{:a}\chi_{:b}\chi_{:a}\chi_{:b}
    +w_{34}\chi_{:aa}\chi_{:bb}+w_{35}\chi_{:ab}\chi_{:ab}$
\smallbreak\qquad\qquad\quad
    $+w_{36}\chi_{:a}\chi_{:a}\chi_{:bb}
     +w_{37}\chi_{:b}\chi_{:aab}\}
    +F_{;m}\{w_{38}\chi_{:a}S_{:a}+w_{39}\chi_{:a}\chi_{:a}L_{cc}$
\smallbreak\qquad\qquad\quad
    $+w_{40}\chi_{:a}\chi_{:b}L_{ab}
     +w_{41}\chi\chi_{:a}\Omega_{am}\}
    +F_{;mm}\{w_{42}\chi_{:a}\chi_{:a}\}$.
\medbreak
The following Theorem represents the main result of this paper:

\begin{theorem}\label{ThmB} We have
$$a_5 (F,D,\cab)=\fract/1/5760/(4\pi)^{-(m-1)/2}
   \ptr({\cal A}_1^1+{\cal A}_5^2+{\cal A}_5^3(\vec w))[\partial M]$$
where the universal constants $\vec w=(w_1,...,w_{42})$ are
given by:\smallbreak
\centerline{\begin{tabular}{|l|l|l|l|l|}\hline
$w_1=-180$ & $w_2=180$ & $w_3=-120$ & $w_4=720$ & $w_5=-\frac {105} 4$\\\hline
$w_6=120$  & $w_7=\frac{105} 4$ & $w_8=-45$ & $w_9=180$ & $w_{10}=-45$\\\hline
$w_{11}=360$ & $w_{12}=45$ & $w_{13}=-180$ & $w_{14}=90$ & $w_{15}=90$\\\hline
$w_{16}=120$ & $w_{17}= 180$ & $w_{18}= 300$ & $w_{19}=-180$ & $w_{20}=-90$\\\hline
$w_{21}=240$&$w_{22}=-30$&$w_{23}=0$&$w_{24}=-60$&$w_{25}=30$\\\hline
$w_{26}=-\frac{675}{32}$&$w_{27}=-\frac{75} 4$&$w_{28}= -\frac{195}{16}$&
 $w_{29}=-\frac{675} 8$&$w_{30}=-330$\\\hline
$w_{31}= -300$&$w_{32}= \frac{15}4$&$w_{33}=\frac{15} 8$&
   $w_{34}=-\frac{15} 4$&$w_{35}=-\frac{105} 2$\\\hline
$w_{36}= -15$&$w_{37}=-\frac{135} 2$&$w_{38}=-210$&$w_{39}=-\frac{165}
{16}$&$w_{40} =- \frac{405} 8$
\\\hline$w_{41}=135$&$w_{42}= -30$&&&\\\hline
\end{tabular}}
\end{theorem}

The remainder of this paper is devoted to the evaluation of these constants; the various
coefficients are determined in the following sections:\medbreak
\centerline{\begin{tabular}{|ll|ll|ll|ll|ll|ll|}\hline
$w_1$& \S\ref{Sectd}&
$w_2$& \S\ref{Sectd}&
$w_3$& \S\ref{Sectd}&
$w_4$& \S\ref{Sectd}&
$w_5$&\S\ref{Sectg}&
$w_6$&\S\ref{Sectc}\\\hline
$w_7$&\S\ref{Sectg}&
$w_8$& \S\ref{Sectd}&
$w_9$& \S\ref{Sectd}&
$w_{10}$& \S\ref{Sectd}&
$w_{11}$& \S\ref{Sectd}&
$w_{12}$&\S\ref{Sectg}\\\hline
$w_{13}$&\S\ref{Sectg}&
$w_{14}$&\S\ref{Sectg}&
$w_{15}$& \S\ref{Sectd}&
$w_{16}$& \S\ref{Sectf}&
$w_{17}$&\S\ref{Sectg}&
$w_{18}$& \S\ref{Sectf}\\\hline
$w_{19}$& \S\ref{Sectc}&
$w_{20}$& \S\ref{Sectd}&
$w_{21}$& \S\ref{Sectf}&
$w_{22}$& \S\ref{Sectc}&
$w_{23}$& \S\ref{Sectf}&
$w_{24}$& \S\ref{Sectf}\\\hline
$w_{25}$& \S\ref{Sectf}&
$w_{26}$& \S\ref{Secte}&
$w_{27}$& \S\ref{Sectf}&
$w_{28}$& \S\ref{Sectf}&
$w_{29}$& \S\ref{Sectf}&
$w_{30}$& \S\ref{Sectf}\\\hline
$w_{31}$& \S\ref{Sectf}&
$w_{32}$& \S\ref{Sectf}&
$w_{33}$& \S\ref{Sectf}&
$w_{34}$& \S\ref{Sectf}&
$w_{35}$& \S\ref{Sectf}&
$w_{36}$& \S\ref{Sectf}\\\hline
$w_{37}$& \S\ref{Sectf}&
$w_{38}$& \S\ref{Sectf}&
$w_{39}$& \S\ref{Sectf}&
$w_{40}$& \S\ref{Sectf}&
$w_{41}$&\S\ref{Sectg}&
$w_{42}$& \S\ref{Sectf}\\\hline\end{tabular}}
\section{Invariance theory}\label{Sectc}

The main result of this section is the following technical result.

\begin{lemma}\label{SectcAa} There exist universal constants $w_i$
so that $$a_5 (F,D,\cab)=\fract/1/5760/(4\pi)^{-(m-1)/2}
   \ptr({\cal A}_1^1+{\cal A}_5^2+{\cal A}_5^3(\vec w))[\partial M].$$
\end{lemma}

Let $N^\nu(F)=F_{;m ... m}$ be the $\nu^{th}$ normal covariant derivative 
of the smearing function $F$. Let $R_{ijkl}$,
$\Omega_{ij}$, and $E$ have order $2$; let $L_{ab}$ and $S$ have order $1$,
and let $\chi$ have order $0$. We increase the order by $1$ for every
explicit covariant derivative. The following Lemma is proved
in \cite{BG90} and summarizes some of the properties of the invariants $a_n$ that we shall need.

\begin{lemma}\label{SectcA}\ \begin{enumerate}
\smallskip\item There exists an invariant local 
formula $a_n^M=a_n^M(x,D)$ on $M$ which is
homogeneous of order $n$ and which vanishes 
if $n$ is odd, and there exist invariant local
formulas $a_{n,\nu}^{\partial M}=a_{n,\nu}^{\partial M}(y,D,\cab)$ on $\partial M$
which are homogeneous of order
$n-\nu-1$ so that
$$a_n(F,D,\cab)=\ptr(Fa_n^M)[M]+\sum_{0\le\nu\le n-1}\ptr(N^\nu(F)
    a_{n,\nu}^{\partial M})[\partial M].$$
\item Let $M=M_1 \times M_2$ for $M_1$ closed. Let
$D=D_1\otimes1+1\otimes D_2$. Then
\begin{eqnarray*}
&&a_n^M(x,D)=\sum_{ p+q=n}a_p^{M_1}(x_1,D_1)a_q^{M_2}(x_2,D_2)\text{ and}\\
&&a_{n,\nu}^{\partial M}(y,D,\cab)=\sum_{p+q=n}a_p^{M_1}(x_1,D_1)
     a_{q,\nu}^{\partial M_2}(y_2,D_2,\cab).\end{eqnarray*}
\item Let $D=D_1\oplus D_2$ over $M$. Then
\begin{eqnarray*}&&a_n^M(x,D)=a_n^M(x,D_1)+a_n^M(x,D_2)\text{ and}\\
  &&a_{n,\nu}^{\partial M}(y,D,\cab)=a_{n,\nu}^{\partial M}(y,D_1,\cab)
  +a_{n,\nu}^{\partial M}(y,D_2,\cab).\end{eqnarray*}
\item We have the following variational formulas:
\begin{eqnarray*}
&&\birdy a_{m-2}(e^{ -2 \epsilon f}F,e^{
-2 \epsilon f}D, \cab )=0\text{ and}\\
&&\birdy a_n(1,e^{-2\epsilon F}D, \cab )=(m-n)a_n(F,D, \cab ).\end{eqnarray*}
\item Let $P:C^\infty(V_1)\rightarrow C^\infty(V_2)$ be an elliptic
complex of Dirac type. Let
$D:=P^*P$ and let
$\hat D:=PP^*$ be the associated operators of Laplace type. 
With suitable boundary conditions, we have:
\begin{eqnarray*}
&&a_m(1,D,\cab)-a_m(1,\hat D,\hat\cab)={\rm index}(P)\text{ and}\\
&&a_n(1,D,\cab)-a_n(1,\hat D,\hat\cab)=0\text{ if }n\ne m.\end{eqnarray*}
\end{enumerate}\end{lemma}

The invariants which do not involve the interaction between
$V_D$ and $V_N$ were determined in \cite{BGV95, K98}. There are no interior
invariants since $5$ is odd. To complete the proof of Lemma \ref{SectcAa}, we
must exhibit a spanning set for the space of boundary invariants which are
homogeneous of order $4$; assertion 2 and 3 of Lemma \ref{SectcA}
show that the constants $w_i$ are dimension free. 
We can eliminate some terms from
consideration:
\begin{lemma}\label{SectcB} Let $X$ be an arbitrary
scalar monomial.\begin{enumerate}
\smallskip\item The following invariants do not appear in $a_5$:
$X\chi \chi_{:a}E_{:a}$,
$\chi \chi_{:a}\chi_{:abb}$,
$\Omega_{am}(S_{:a}\chi+\chi S_{:a})$,
$\chi\ca\cb X_{ab}$,
$\Omega_{am}S_{:a}$,
$\fm\chi\ca S_{:a}$,
$(\chi_{:aa}\chi-\chi\chi_{:aa})E$,
$X_{ij}\chi_{:a}\Omega_{kl}$, and
$\chi_{:a}\Omega_{jk;i}$.
\smallskip\item The following invariants are linear combinations of the invariants defining ${\cal
A}_5^3$:
$\chi \chi_{:aa}\chi_{:bb}$,
$\chi \chi_{:ab} \chi_{:ab}$,
$\chi \chi_{:aa}\chi\chi_{:bb}$,
$\chi \chi_{:ab}\chi\chi_{:ab}$,
$\chi \chi_{:ab} \ca\cb$,\newline
$\chi \chi_{:ab} \cb \ca$, and
$\chi \chi_{:aa} \cb\cb$.
 
\smallskip\item We have the following relations:
   $\ptr(\chi\ca\cb\ca\cb)=0$,\newline
   $\ptr(\chi\ca X_a)=0$,
   $(\ptr\chi\ca\ca X)=0$,
   $\ptr(\chi_{:ab}X_{ab})=0$, and\newline
  $\ptr(\ca X_a)=0$.
\end{enumerate}\end{lemma}

\proof We use arguments of \cite{BG92} to prove the first assertion. Let the
smearing function $F$ be real. Suppose that the bundle $V$, the operator $D$,
and the endomorphisms
$S$ and
$\chi$ are real. It then follows that $\ptr(Fe^{-tD_\cab})$ is real.
Thus all the coefficients must be real. Suppose next that $V$ is complex,
that $\nabla$ is Hermitian, and that $E$, $S$, and $\chi$ are self-adjoint.
Then the operator $D_\cab$ is self-adjoint and again $\ptr(Fe^{-tD_\cab})$ is
real. The curvature tensor $\Omega_{ij}$ is anti-Hermitian as it is a
commutator of two covariant derivatives. The terms in the first assertion
have purely imaginary trace and thus do will not
appear in the formula for $a_5^{\partial M}$.

Let $\{A,B\}:=AB+BA$ be the anti-commutator. We differentiate the
identity $\chi^2=1$ to see
\beq
\{ \chi , \ca \} =0 , \quad \{ \chi_{:ab}, \chi \} +
        \{ \ca , \cb \} =0, \nonumber\\
\{\chi_{:aab} , \chi \} +\{ \chi_{:aa}, \cb \} +         
            2\{ \chi_{:ab} , \ca \} =0.\label{mix3}\nn
\eeq

We use the Ricci identity to see that, at the cost of getting
additional terms involving $\Omega$, the fourth derivative of $\chi$ can
be reduced to terms with lower derivatives. Thus 
$$\{\chi_{:aabb}, \chi \} + 
   \mbox{(fewer derivatives in }\chi ) =0.$$
We differentiate the relation $\chi S=S\chi=S$ to see that:
\begin{eqnarray}
&&S\ca +S_{:a} \chi = S_{:a} = \ca S +\chi S_{:a}\nonumber\\
&&S_{:ab}=S\chi_{:ab} +S_{:a} \cb +S_{:b} \ca +S_{:ab} \chi\label{mix4}\\
&&\qquad=\chi_{:ab} S +\ca S_{:b} +\cb S_{:a} +\chi S_{:ab}.\nonumber
\end{eqnarray}
We also observe that:
\begin{eqnarray*}
&&\ptr(\ca \cb \chi_{:ab}) = \frac 1 2 \ptr(\ca \cb \chi \{\chi ,
\chi_{:ab} \}) = \frac 1 2 \ptr(\chi \ca\ca\cb\cb ) \nn\\
&&\ptr (\ca\ca\chi_{:bb} ) = -\ptr (\chi \ca\ca\cb\cb ) .\nn
\end{eqnarray*}
We use Equation (\ref{mix4}) to see that we do not need a $\chi$, a $\ca$ nor
a 
$\chi_{:aa}$ touching an $S$. We use the identities $\ptr (\chi \ca S_{:b} L_{ab} )=0$ and
\beq
\ptr ( S\chi_{:ab} L_{ab} )& =& -\ptr(S\ca\cb L_{ab} ) = 
         -\ptr (S_{:a} \cb L_{ab} )\nn
\eeq
together with Equation
(\ref{mix4}) to see that $\ptr(\ca S)=0$, 
$\ptr (\chi S_{:b} )= \ptr (S_{:b})$,
\beq
\ptr (\chi S_{:a} \chi S_{:a} ) &=& \ptr (2\chi S_{:a} S_{:a} -
            S_{:a} S_{:a} ),\text{ and } \nn\\
\ptr (\chi S_{:ab} L_{ab} ) &=& 
      \ptr (S_{:ab} L_{ab} -S\chi_{:ab} L_{ab} -2S_{:a} \chi_{:b} L_{ab} )
      \nn\\
    &=& \ptr (S_{:ab} L_{ab} -S_{:a} \cb L_{ab} ).\ \qedbox\nn
\eeq

In addition to the relations of Lemma \ref{SectcB}, we note that terms
obtained by commuting the order of derivatives are controlled by the
Ricci identity. 

\medbreak\noindent{\bf Proof of Lemma \ref{SectcAa}:} \rm We have
eliminated most of the additional possible terms which might otherwise be
thought to appear in the local formula for $a_5$. 
Note that terms of the form $\chi \chi_{:ab} X_{ab} $ 
are controlled by $\ca \cb X_{ab}$. There are, however, a few
remaining terms to be eliminated. We order these terms by
length and complete the proof by eliminating these terms.\begin{enumerate}
\item (Terms of length 2). 
    As $\Omega_{ab}$ is antisymmetric, we have
   $\ptr (\chi_{:ab}\Omega_{ab}) =0$. 
\item (Terms of length 3). We use the Ricci identity to see that 
\beq
\chi \chi_{:ab} \Omega_{ab} = \frac 1 2 \chi (\chi_{:ab} -\chi_{:ba} ) 
\Omega_{ab} = \frac 1 2 \chi [\chi , \Omega_{ab} ] \Omega_{ab}. \nn
\eeq
The terms on the right hand side are controlled and thus
we may omit $\ptr(\chi\chi_{:ab}\Omega)$ from our list of
invariants. A similar argument shows we may omit
$\ptr(\chi_{:ab} \chi \Omega_{ab})$ from our list of invariants. 
We use Lemma
\ref{SectcB} to exclude the invariant $\chi\chi_{:aa}E-\chi_{:aa} \chi E$. We
control the invariant $\chi\chi_{:aa}E+\chi_{:aa} \chi E$ by
$\ca
\ca E$. 
\item(Terms of length 4). By Equation (\ref{mix3}),
$\ptr (\chi \chi_{:ab} \chi \Omega_{ab}) =0$. \qedbox\end{enumerate}

The following relations follow from Lemma \ref{EQGc} and from the product
formula of assertion 2 of Lemma \ref{SectcA}.

\begin{lemma}\label{SectcX} We have
$w_3+w_4=600$, $w_{19}=-180$, $w_{22}=-30$, $w_5+w_7=0$, and $w_6=120$.
\end{lemma}
\section{Two dimensional calculations}\label{Sectd}

Let $D$ be an operator of Laplace type on the trivial vector bundle over the cylinder
$M:=S^1\times [0,1]$, which we endow with the standard flat metric. We specialize the formula for 
$a_5$ to the situation at hand. 
We set $L=0$, $R=0$,
$F=1$, $m=2$, and $a=b=c=1$.
This yields the expression: 
\begin{eqnarray}
    &&a_5(1,D,\cab)=5760^{-1}(4\pi)^{-1/2}\ptr\{360\chi 
       E_{;22}+1440 E_{;2}S+720\chi E^2
     \nonumber
      \\&&\qquad+240\chi  E_{:11}
       +960SS_{:11}
       +2880ES^2+1440S^4+w_1E^2\nonumber\\
    &&\qquad+w_2\chi E\chi E+w_3S_{:1}S_{:1}
     +w_4\chi S_{:1}S_{:1}
    +w_8\Omega_{12}\Omega_{12}\nonumber\\
    &&\qquad+w_9\chi\Omega_{12}\Omega_{12}
    +w_{10}\chi\Omega_{12}\chi\Omega_{12}
    +w_{11}(\Omega_{12}\chi S_{:1}-\Omega_{12}S_{:1}\chi)\nonumber\\
    &&\qquad+w_{15}\chi\chi_{:1}\Omega_{12;2}
      +w_{18}\chi_{:1} E_{:1}
    +w_{19}\chi_{:1}\chi_{:1}E+w_{20}\chi\chi_{:1}\chi_{:1}E\nonumber\\
    &&\qquad+w_{21}\chi_{:11}E
    +w_{32}\chi_{:1}\chi_{:1}\chi_{:1}\chi_{:1}
    +w_{33}\chi_{:1}\chi_{:1}\chi_{:1}\chi_{:1}
    +w_{34}\chi_{:11}\chi_{:11}\nonumber\\
    &&\qquad+w_{35}\chi_{:11}\chi_{:11}
    +w_{36}\chi_{:1}\chi_{:1}\chi_{:11}
     +w_{37}\chi_{:1}\chi_{:111}\}[\partial M].\nonumber\end{eqnarray}
We use the relation $w_3+w_4=600$ to see that
\begin{eqnarray*}&&\ptr(960SS_{:11}+w_3S_{:1}S_{:1}
+w_4\chi S_{:1}S_{:1})[\partial M]\\
&=&\ptr(-360S_{:1}S_{:1}-2w_4\Pi_-S_{:1}S_{:1})[\partial M].\end{eqnarray*}
We integrate by parts to see
$f_{:1}g[\partial M]=-fg_{:1}[\partial M]$.
We use this relation to see $\ptr(\chi_{:1}\chi_{:1}\chi_{:11})[\partial M]=0$
and to simplify the expression for $a_5$. This shows:
\begin{eqnarray}
    &&a_5(1,D,\cab)=5760^{-1}(4\pi)^{-1/2}\ptr\{360\chi E_{;22}
      +1440 E_{;2}S+720\chi E^2
     \nonumber
      \\&&\qquad
       +2880ES^2+1440S^4+w_1E^2+w_2\chi E\chi E
      -360S_{:1}S_{:1}\nonumber\\
    &&\qquad
     -2w_4\Pi_- S_{:1}S_{:1}
    +w_8\Omega_{12}\Omega_{12}+w_9\chi\Omega_{12}\Omega_{12}
    +w_{10}\chi\Omega_{12}\chi\Omega_{12}\nonumber\\
    &&\qquad
    +w_{11}(\Omega_{12}\chi S_{:1}-\Omega_{12}S_{:1}\chi)
     +w_{15}\chi\chi_{:1}\Omega_{12;2}
    +w_{19}\chi_{:1}\chi_{:1}E\nonumber\\
    &&\qquad+w_{20}\chi\chi_{:1}\chi_{:1}E
    +(240-w_{18}+w_{21})\chi_{:11}E\nonumber\\
    &&\qquad
    +(w_{32}+w_{33})\chi_{:1}\chi_{:1}\chi_{:1}\chi_{:1}
    +(w_{34}+w_{35}-w_{37})\chi_{:11}\chi_{:11}\}[\partial M].\nonumber\end{eqnarray}

\begin{lemma}\label{SectdA} We have
\begin{eqnarray*}
\begin{array}{|l|l|l|l|}\hline
w_1 = -180&w_2 = 180&w_3 = -120&
w_4 = 720\\\hline
w_8 = -45&w_9 = 180&
w_{10} = -45&w_{11} = 360\\\hline w_{15} = 90&
w_{19}= -180&w_{20} = -90&w_{21} -w_{18} = -60\\\hline
\end{array}\end{eqnarray*}\end{lemma}

\proof  We apply the local index formula of Lemma \ref{SectcA} (5). Let
$i$, $j$, and $k$ be real skew-adjoint
$ 4 \times 4$ matrices satisfying
the quaternion relations. Let
\begin{eqnarray*}
&& A:=a_{ 0}+ia_1+ja_{ 2}+ka_{
3} \quad {\rm for}\quad
   \vec a \in  \preal  \oplus \sqrt{-1} \preal^{
3} \nonumber \\
&& B:=b_{ 0}+ib_1+jb_{ 2}+kb_{
3}\quad {\rm for}\quad
   \vec b \in \sqrt{-1} \preal  \oplus  \preal^{
3} 
\end{eqnarray*}
be matrix valued functions on
$     M$. Then
$ A^{ *}=A$ and
$ B^{ *}=-B$. Let
$$
 P:=i \partial_1+j \partial_{
2}+A+B \quad {\rm and} \quad
 P^{ *}:=i \partial_1+j \partial_{
2}+A-B$$
be operators of Dirac type and let
$ D:=P^{ *}P$ and
$   \hat D:=P\, P^{ *}$ be the associated
operators of Laplace type. Let
$  \chi :=\sqrt{-1}i;$ we note that
$  \chi^2=1$,
$  \chi i=i \chi $, and
$  \chi j=-j \chi $. Let
$  \Pi_{  \pm }:= {\textstyle{1\over2}} (1 \pm  \chi )$
be the complementary projections on the
$  \pm 1$ eigenspaces of
$  \chi $. We define admissible boundary operators $\cab$ and $\hat\cab$ of
mixed type which make $D$ and $\hat D$ self-adjoint:
$$ \cab \phi :=( \Pi_{ -}\phi ) |_{
\partial M} \oplus ( \Pi_{ -}P\phi ) |_{
\partial M },\text{ and }   \hat  \cab \phi :=( \Pi_{ -}\phi ) |_{
\partial M} \oplus ( \Pi_{ -}P^{
*}\phi ) |_{  \partial M}.$$

Since
$     P$ intertwines
$     D$ and
$   \hat D$ and since the index of this complex is zero, 
$$a_n(1,D, \cab )=a_n(1,  \hat D,  \hat  \cab ).$$
If we interchange
$     B$ and
$ -    B$, then we interchange the roles of
$     P$ and
$ P^{ *}$ and the roles of
$     D$ and
$   \hat D$. Thus the terms of odd degree
in
$     B$ must vanish in
$ a_n(1,D, \cab )$. Let
$ \dot A= \partial_1A$,
$ \dot B= \partial_1B$,
$   \tilde A= \partial_{ 2}A$, and
$   \tilde B= \partial_{ 2}B$. We first
study the terms which are bilinear in the
jets of
$     A$ and
$     B$. These terms change sign if we interchange
the roles of
$     D$ and
$   \hat D$ and hence their coefficient in
$ a_{ 5}$ must be zero. Let
$  {\cal J} :=2\ptr(I)$ and
$  {\cal K} :=2\sqrt{-1}\ptr(I)$. We use
Lemmas A.1-A.3 to obtain the following system
of Equations.

\medbreak\noindent
\qquad
$ 0=2  w _8+2  w_{10}+2  w_{15}$ \hfill
$ (  \tilde a_{ 0}\dot b_{ 3} {\cal J} ) $

\par\noindent
\qquad
$ 0=-2  w _8+2  w_{10}$
\hfill
$ (\dot a_{ 0}  \tilde b_{ 3} {\cal J} ) $

\par\noindent
\qquad
$ 0=-2  w_9+2 (240-w_{18}+w_{21})$
\hfill
$ (  \tilde a_1\dot b_{ 3} {\cal K} ) $

\par\noindent
\qquad
$ 0=-2  w_{2}-2  w_{1}$
\hfill
$ (  \tilde a_{ 3}\dot b_{ 0} {\cal J} ) $

\par\noindent
\qquad
$ 0=-2  w_{2}+2  w_{1}+ {\textstyle{1\over2}} 1440$
\hfill
$ (\dot a_{ 3}  \tilde b_{ 0} {\cal J} ) $

\par\noindent
\qquad
$ 0=2 (240-w_{18}+w_{21})-360$ \hfill
$ (\dot a_{ 3}\dot b_{ 2} {\cal K} ) $

\par\noindent
\qquad
$ 0=4  w _8+4  w_{10}+1440-12  w_{15}$ \hfill
$ (  \tilde a_{ 0}a_{ 0}b_{ 2} {\cal J} ) $

\medbreak\noindent
This implies that:
\begin{eqnarray*}
&&   w _8=-45,\
   w_9=180,\
   w_{10}=-45,\
   w_{2}=180, 
\nonumber \\
&&   w_{1}=-180,\
  (240-w_{18}+w_{21})=180,\
   w_{15}=90.
\end{eqnarray*}
Next we set
$ a_{ 0}=a_1=a_{ 2}=0$, we let
$a_{ 3}$ be a constant,
and we assume
$ B=B(x_1)$. This yields the following
system of Equations we use to complete the
proof of Lemma \ref{SectdA}.
\smallbreak\hfill{\begin{tabular}{lr}
$ 0=4  w_{20}-6 (240-w_{18}+w_{21})+2 \cdot 720$\qquad\qquad\qquad&
$ (\dot b_{ 0}b_{ 3}^2 {\cal K} ,b_{
0}\dot b_{ 3}b_{ 3} {\cal K} ) $\\
$ 0=-8  w_{20} -2w_4+8 (240-w_{18}+w_{21})$\\
$\phantom{0=}+2(-360)-2880+2 \cdot 1440$&
$ (a_{ 3}b_{ 2}^{ 3} {\cal K} ) $\\
$ 0=4  w _8+4  w_{10}-4  w_{15}$\\
$\phantom{0=}-2(-4  w _8+4  w_{10}+4  w_{15}+4  w_{19})$&
$ (\dot b_{ 3}b_{ 3}b_1 {\cal J} ,\dot b_1b_{ 3}^2 {\cal J} ) $\\
$ 0=4  w_{19}+ {\textstyle{1\over2}} 2880-2  w_{11}.\ \qedbox$&
$ (\dot b_1b_{ 2}^2 {\cal J} ) $\end{tabular}}

\section{Calculations on the ball}\label{Secte}

In this section, we compare the results of calculations on the ball performed
in \cite{DAKB96,stukla} with the formula from 
Theorem \ref{ThmB} to establish the
following result:

\begin{lemma}\label{SecteA} We have the following relations:
\begin{eqnarray*}
-135&=&2 w_{27} +4 w_{33} +2 w_{35}\\
705&=&16 w_{28} +16 w_{29} -4 w_{31} +16 w_{32} -16 w_{33} 
+16 w_{34} -16 w_{37}\\
1725&=&2w_3 +32 w_{26} -8 w_{30}\\
-675&=&32 w_{26}\\
1935&=&16 w_{28} +16 w_{29} -8 w_{30} +32 w_{32} +32 w_{34} 
+16 w_{35} +32 w_{36} -32 w_{37}\\
585&=&4 w_{27} -2 w_{31} +8 w_{32} +16 w_{33} +8 w_{34} 
+12 w_{35} -8 w_{36} -8 w_{37}.\end{eqnarray*}\end{lemma}

\proof We start with the Dirac spinors. Let $\gamma_i\in M_{2^{[m/2]}}$ be
the Dirac gamma matrices; these satisfy the Clifford
commutation relation $\gamma_j \gamma_k + \gamma_k \gamma_j = -2\delta_{kj}$.
Let $P=\gamma^\mu \nabla_\mu$ be an operator of Dirac type. We assume that
the connection is compatible, in other words $\nabla\gamma=0$, or in
index notation, $[\nabla_\mu
,\gamma_\nu ]=0$. For
$m$ even, let
$$
\Gamma^5:=(\sqrt{-1})^{m/2}\gamma_1\dots\gamma_m.$$
We then have
$(\Gamma^5)^2=1$ and $\Gamma^5\gamma_j+\gamma_j\Gamma^5=0$.
We refer to \cite{BG92} for further information concerning the spectral geometry
of the Dirac operator with local boundary conditions. We use formulas from
\cite{BG92}. 
Consider the boundary value problem for the
corresponding Laplacian $D=P^2$. We set
$$
\chi: = -\Gamma^5 \gamma_m\text{ and }S:=-\textstyle\frac 12 L_{aa} \Pi_+
$$
to define admissible boundary conditions. The endomorphism $E$, the Riemann curvature and $\Omega$
are zero. Since $L_{ab:c}=0$ and $L_{ab}=\delta_{ab}$, we have:
$$\chi_{:a} =L_{ac}\Gamma^5 \gamma_c,\ 
\chi_{:ab}=-L_{ac}L_{bc}\chi,\ 
S_{:a}=-\textstyle{\frac 14} L_{bb} \chi_{:a},\ 
S_{:ab}=\textstyle{\frac 14} L_{dd} L_{ac}L_{bc}\chi.$$
Dowker, Apps, Kirsten and Bordag showed in \cite{DAKB96} that:
$$
\begin{array}{rl}
a_5(1,D,\cab_S)&=\displaystyle \frac{2^{-m+1}2^{m/2}\sqrt{\pi}}
{\Gamma({\frac{m}{2}})} 
     \left(-\frac{29(m-1)}{2560} +\frac{91 (m-1)^2}{30720}\right.  \\
&\left.\qquad\qquad\qquad\qquad+\displaystyle
\frac{11 (m-1)^3}{3840} -\frac{89 (m-1)^4} {122880}\right).
\end{array}
$$
We have
${\rm Vol}(S^{m-1})=2\pi^{m/2}/{\Gamma({\frac{m}{2}})}$.
We compare coefficients
of the powers of
$m-1$ in the above equation with the general formula of Theorem \ref{ThmB}
to establish the first three equations of Lemma \ref{SecteA}.

Next we study vector fields (or 1-forms) on the unit ball
with absolute boundary conditions. For these boundary
conditions, $\chi$ can be viewed as a matrix acting
in the tangent space to $M$:
$$\chi_{mm} = -1, \quad \chi_{ba}=\delta_{ba},\text{ and } \quad
S= -\Pi_+ .$$
As above, $E=0$, $R=0$, and $\Omega=0$. We have that:
\begin{eqnarray*}
&&(\chi_{:a})_{mb}=(\chi_{:a})_{bm}=-2\delta_{ab}
\nonumber \\
&&(\chi_{:ab})_{mm}=4\delta_{ab}\ ,\qquad
(\chi_{:ab})_{cd}=-2(\delta_{cd}\delta_{ad} +
\delta_{ac}\delta_{bd}).
\end{eqnarray*}

We use \cite{stukla}
(see also \cite{ELV,V95-2}) to see the 
following relationship from which the
final assertions of Lemma \ref{SecteA} follow:
\begin{eqnarray*}\displaystyle
\frac{(4\pi )^{(m-1)/2}}{{\rm Vol}\, S^{m-1}} a_5 &=& \displaystyle
{\frac{1309\,(m-1)}{15360}} - {\frac{6313\,{(m-1)^2}}{36864}} + 
   {\frac{6359\,{(m-1)^3}}{46080}}\\ &&\displaystyle - {\frac{26587\,{(m-1)^4}}{737280}} + 
   {\frac{2041\,{(m-1)^5}}{737280}}.\ \qedbox\end{eqnarray*}
\section{Conformal relations}\label{Sectf}

\begin{lemma}\label{SectfA}We have the following relations:
\smallbreak\centerline{\begin{tabular}{|l|l|l|l|}\hline
      $w_{16} = 120$&
      $w_{18}= 300 $&
      $w_{21}=240$&
      $w_{23}=0$\\\hline
      $w_{24}= -60 $&
      $w_{25}= 30$&
      $w_{26}= -\frac{675}{32}$&
      $w_{27}= -\frac{75} 4$\\\hline
      $w_{28}= -\frac{195}{16}$&
      $w_{29}=-\frac{675} 8$&
      $w_{30} = -330$&
      $w_{31}=-300$\\\hline
      $w_{32}= \frac{15} 4$&
      $w_{33}= \frac{15} 8$&
      $w_{34}= -\frac{15} 4$&
      $w_{35} = -\frac{105} 2$\\\hline
      $w_{36}= -15$&
      $w_{37}=-\frac{135} 2$&
      $w_{38}=-210$&
      $w_{39}= -\frac{165}{16}$\\\hline
      $w_{40} =- \frac{405} 8$&
      $w_{42} = -30$&&\\\hline
\end{tabular}}\end{lemma}

\proof We use Lemma \ref{SectcA} (4). Consider the
variations
$ D( \epsilon )=e^{ -2 \epsilon f}D$,
$g( \epsilon )=e^{ 2 \epsilon f}g$, and
$F( \epsilon )=e^{ -2 \epsilon f}F$.
We set
$$ 
S( \epsilon )=e^{ - \epsilon f}  \Pi_+
\{  \omega_{ m}(0)- \omega_{
m}( \epsilon )+S \} .
$$ 
to fix the boundary conditions. Let
$ e_{ i}$ be an orthonormal frame for
the tangent and cotangent bundles of
$  M$ with respect to the reference metric
$  g(0)$. Let
$ e_{ i}( \epsilon )=e^{ - \epsilon f}e_{
i}$ and
$ e^{ i}( \epsilon )=e^{  \epsilon f}e^{
i}$ be the corresponding frames for the metric
$ g( \epsilon )$. We remark that contraction
and differentiation  do not commute;
$$
  \birdy ( \Phi_{ ii})=( \birdy  \Phi )_{
ii}-2f \Phi_{ ii}
$$
for example. Although the Christoffel symbols
$  \Gamma $ are not tensorial, their variation
is tensorial. We have the following relations; a more extensive list is given in \cite{BGV95}.  Conformal
variations of the new invariants are listed in
Appendix B.
\begin{eqnarray}
&&( \birdy  \Gamma )_{ ij }{}^{ k}= \delta_{
ik}f_{ ;j}+ \delta_{ jk}f_{ ;i}- \delta_{
ij}f_{ ;k}, \nonumber \\
&&( \birdy L)_{ ab}=- \delta_{ ab}f_{
;m}-fL_{ ab}, \nonumber \\
&&  \birdy (S)=-fS+ \Pi_+{\textstyle{1\over2}} (m-2)f_{
;m} , \nonumber \\
&&  \birdy (E)=-2fE+ {\textstyle{1\over2}} (m-2)f_{
;ii},\text{ and} \nonumber \\
&& ( \birdy R)_{ ij k l }= \delta_{
ik}f_{ ;j l }+ \delta_{ j l }f_{
;ik}- \delta_{ i \ell }f_{ ;jk}- \delta_{
jk}f_{ ;il}+2fR_{ ij kl}. \nonumber
\end{eqnarray} 

The following relations are obtained from Lemma
\ref{SectcA} (4) by comparing coefficients before various terms (listed on the left).
Lemma \ref{SectfA} follows from these relations.
\beq
\begin{array}{ll}
\underline{\mbox{Term}} & \underline{\mbox{Coefficient}} \\
F\chi_{:a}E_{:a} & 0 = 2w_{18}-(m-3)w_{21}-360+1440(m-2)\\
  & \phantom{0=}-240(m-3)-
                  960(m-1) +240  \\
F_{;m} \chi_{:a}S_{:a} & 0=(m-1)w_{30}-\frac 1 2 (m-2) w_3 +
    480 (m-2)  \\ & \phantom{0=}+w_{31} +(m-5) w_{38} \\
F_{;mm} \chi_{:a} \chi_{:a} & 0= \frac 1 2 (m-2) w_{19} -2(m-1) w_{22} 
    -(m-1) w_{23} \\
      &\phantom{0=} -w_{24} -w_{25} -(m-5) w_{42}\\
F_{;m} L_{bb} \chi_{:a}\chi_{:a} &  0= \frac 1 2 (m-2) w_{19} -2(m-1) w_{22}
    -w_{23} -w_{24} +w_{29} \\
& \phantom{0=}+2(m-1) w_{26} +2w_{28}  
   -\frac 1 4 (m-2) w_{30}\\
 & \phantom{0=} 
          +(m-5) w_{39}\\
F\chi_{:aab}\chi_{:b}  & 0= (m-2) w_{19} -4(m-1) w_{22} -2w_{23}
-
        2(m-1) w_{24} \\
 & \phantom{0=} -2w_{25} -2(m-3) w_{34} 
     +2w_{35} +(m-1) w_{37} \\
F\chi_{:ab}\chi_{:ab} & 0= (m-2) w_{19} -4(m-1) w_{22} -2w_{23}\\
 & \phantom{0=} -mw_{24} 
            -w_{25} +4w_{35}  \\
F_{;m} \chi_{:a} \chi_{:b} L_{ab}  & 0= -\frac 1 4 (m-2) w_{31} 
     -(m-2) w_{24} -w_{25} +2 w_{27} \\
    & \phantom{0=} +(m-1) w_{29} +(m-5) w_{40} \\
F\chi_{:aa}\chi_{:bb} & 0= -(m-2) w_{24} -w_{25} -2(m-3) w_{34} -2w_{35} \\
 & \phantom{0=}
          +(m-1) w_{37}\\
F_{;m} \chi \chi_{:a} \Omega_{am} & 0= +\frac 1 2 (m-2) w_{11} 
             -(m-1) w_{12} -2w_{15} -w_{17} \\
 & \phantom{0=} -(m-5) w_{41}\\
F\chi_{:a} \chi_{:b} \Omega_{ab} & 0= w_{15} +(m-5) w_{16} +4(m-2) w_{19} 
      -16 (m-1) w_{22} \\
& \phantom{0=} -8w_{23}-(2m+4) w_{24} 
         -2w_{25} +12 w_{35} \\
F\chi\chi_{:a} \Omega_{ab:b} & 0= -w_{15} -(m-5) w_{16} 
          -2(m-2) w_{19} +8(m-1) w_{22} \\
  & \phantom{0=} +4w_{23} +4w_{24} -4w_{35} \\
F\Omega_{ab} \Omega_{ab} & 0= -\frac 1 2 w_{15} -\frac 1 2 (m-5) w_{16} 
        -(m-2) w_{24} -w_{25} +2 w_{35} \\
f_{;m}F_{;m} \chi_{:a} \chi_{:a} & 0=w_{40} +5w_{42} + 6w_{39} -
\frac 54 w_{38}.\ \qedbox
\end{array} 
\nonumber
\eeq                           

\section{Two more index theorem examples}\label{Sectg}
The only remaining universal constants to be determined are the coefficients of
the following terms:
\beq
P &:=& \textstyle{\frac 1 2 }(w_5 -w_7) F (\Om_{ab} \Om_{ab} -\chi \Om_{ab} \chi \Om_{ab}
) +w_{12} F\chi \ca \Om_{am} L_{cc} 
     \nonumber\\& & +w_{13} F \ca \cb \Om_{ab}
+w_{14} F \chi \ca\cb \Om_{ab} \nonumber\\&& +w_{17} F \chi \ca \Om_{bm} L_{ab} 
+w_{41} F_{;m} \chi \ca \Om_{am} .\label{unknown}
\eeq

\begin{lemma}\label{Sectga} We have \smallbreak\centerline{
\begin{tabular}{|l|l|l|}\hline
$-\frac 1 2 (w_5-w_7) = \frac{105\vphantom{{}^A}} 4\vphantom{{}_{{}_A}}$&
$w_{17} = -180$&
$w_{12} = -45$\\\hline
$w_{13} = -180$&$w_{14}=90$&\\\hline\end{tabular}}
\end{lemma}
Let $M$ be the generalized cylinder 
$M:=S^1\times S^1....\times S^1\times[0,1]$ for $m$ even
with the conformally flat metric 
$$ds^2=e^{2f}(dx_1^2+...+dx_m^2).$$
Let $\gai$ be skew-adjoint matrices satisfying the Clifford
relations. Let
$$A:=e^{-f}(\gai \partial_i)\text{ and }
A^*:=e^{-mf}(\gai \partial_i)e^{(m-1)f}.$$
Let $D^{[0]}:=A^*A$ and $D^{[1]}:=AA^*$
be the associated Laplacians. We set 
\begin{eqnarray*}
&&\Gamma^5:=(\sqrt{-1})^{m/2}\gamma_1\dots\gamma_m,\ 
  \chi:=-\Gamma^5\gamma_m\\
&&  \cab^{[0]} \phi :=( \Pi_{ -}\phi ) |_{
\partial M} \oplus ( \Pi_{ -}A\phi ) |_{
\partial M },\text{ and} \nonumber \\
&&  \cab^{[1]} \phi :=( \Pi_{ -}\phi ) |_{
\partial M} \oplus ( \Pi_{ -} A^{
*}\phi ) |_{  \partial M}.
\end{eqnarray*}
We use the local index formula of Lemma \ref{SectcA} (5) to see
that
$$
a_5 (1,D^{[0]},\cab^{[0]})=a_5(1,D^{[1]},\cab^{[1]}).
$$
We use this relationship to derive certain additional relations among
the coefficients appearing in Equation (\ref{unknown}). Let a comma denote ordinary
partial differentiation.
For simplicity we put $f\vert_{\partial M}=0$. This implies also
$f_{,a}\vert_{\partial M}=0$, but, in general,
$f_{,ma}\vert_{\partial M}\neq 0$. 
We compute:
\beq
D^{[1]} &=& e^{-2f} \big\{ -\partial_i^2 -\textstyle\frac 12 m 
 f_{,i} [\gai ,\gaj ]\pa_j +(2-m) f_{,i} \pa_i \nonumber\\ & &  
\qquad+(m-1) f_{,i} f_{,i}
         -(m-1) f_{,ii}\big \},\nonumber\\
D^{[0]} &=& e^{-2f} \{ -\partial_i^2 +\textstyle\frac 12 (m-2) 
    f_{,i} [\gai ,\gaj ] \pa_j +(2-m) f_{,i} \pa_i \},\nn\\
\omega_i^{[1]} &=& -{\textstyle\textstyle{\frac 1 4}} m [\gai ,\gaj ]f_{,j}, \nn\\
\Om_{ij}^{[1]} &=& -\textstyle{\frac 1 4} m \left\{ f_{,ki}[\gaj ,\gak ]
     -f_{,kj} [\gai ,\gak \right\} \nn\\
 && +\textstyle{\frac 1 4}m^2 \left\{ -f_{,j}f_{,k} [\gai ,\gak ] +
     f_{,i}f_{,k}[\gaj ,\gak ] +f_{,k}f_{,k}[\gai ,\gaj ]
     \right\},\nn\\
E^{[1]} &=& e^{-2f} (m-1)\{ f_{,ii} +\textstyle{\frac 1 4}f_{,i} f_{,i} 
      (m^2-4)\},\nn\\
S^{[1]} &=& (m-1) f_{,m} \Pi_+, \nn\\
\omega_i^{[0]} &=& \textstyle{\frac 1 4} (m-2) [\gai ,\gaj ] f_{,j},\nn\\
\Om_{ij}^{[0]} &=& \textstyle{\frac 1 4}(m-2) \left\{ f_{,ki}[\gaj ,\gak ]
     -f_{,kj} [\gai ,\gak \right\} \nn\\
 && +\textstyle{\frac 1 4}(m-2)^2 \left\{ -f_{,j}f_{,k} [\gai ,\gak ] +
     f_{,i}f_{,k}[\gaj ,\gak ] +f_{,k}f_{,k}[\gai ,\gaj ]
     \right\},\nn\\
E^{[0]} &=& \textstyle{\frac 1 4}e^{-2f} (m-1)(m-2)^2 
  f_{,i} f_{,i},\text{ and}
     \nn\\
S^{[0]} &=& 0 .\nonumber
\eeq
The metric and hence the Riemann tensor is the same for both operators:
\beq
{R^i}_{jkl}&=&f_{,j}f_{,k}\delta_{il} +f_{,i}f_{,l}\delta_{jk}
-f_{,j}f_{,l}\delta_{ik} -f_{,i}f_{,k}\delta_{jl}\nn\\
&&\quad+f_{,p}f_{,p}(\delta_{jl}\delta_{ik}
-\delta_{jk}\delta_{il}).
\nn\eeq

We can now compute the invariants $a_5 (1,D^{[i]},\cab^{[i]})$.
In Equation (\ref{unknown}), the terms $\frac1 2 (w_5-w_7),w_{12},w_{13}$ and
$w_{17}$ contribute to $a_5$ in this setting. The invariants that  
appear in these contributions are $f_{,am}f_{,am}$, $f_{,mm} f_{,m}^2$, $f_{,m}^4$, and 
thus only terms of these types need to be kept during the calculation.
In Appendix C we have listed all contributions appearing in $a_5
(1,D^{[1]},\cab^{[1]})-a_5 (1,D^{[0]},\cab^{[0]})$. We therefore have the relations:
\beq
\begin{array}{ll}
\underline{\mbox{Term}} & \underline{\mbox{Coefficient}} \\
f_{,am}f_{,am} & 0=105+4 (1/2) (w_5 -w_7) \\
f_{,m}^2 f_{,mm} & 0 = 135 -w_{12} +w_{17} +m(45 +w_{12}) \\
f_{,m}^4 & 0 = 180 +w_{13}. 
\end{array}
\eeq
We use Lemma \ref{SectcX} to determine $w_5$ and $w_7$;
$w_{41}$ is now determined by the conformal relation
of the previous section. This determines all the coefficients but $w_{14}$.

Let $M$ be as in Lemma 6.1, let
 $A=\gai (\partial_i +\chi f_i)$, let $f_m$ be imaginary,
and $f_a$ be real.  It is obvious that $a_5(1,A^*A)=a_5(1,AA^*)$.
Since 
$A^* =\gai (\partial_i -\chi f_i)$,
the heat kernel for $AA^*$ is obtained from that for
$A^*A$ by changing sign before $f_i$. Therefore,
all coefficients before odd powers of $f_i$ must vanish. 
Thus the coefficient before $f_m^2 f_{a,a}$ in
$a_5(1,A^*A)$ is zero.
As above, $S=0$. We compute:
\begin{eqnarray*}
&&\omega_a=\frac 12 [\gab ,\gaa ] \chi f_b -\gaa \gam \chi f_m
\nonumber \\
&&\omega_m =\chi f_m \nonumber \\
&&\Omega_{am}=f_{m,a}\chi -2\gaa\gam f_m^2 \nonumber \\
&&\Omega_{ab}=-\frac 12 \left( f_{c,a} [\gab ,\gac ] -
f_{c,b}[\gaa ,\gac ] \right) \chi -( f_{m,b} \gaa -f_{m,a}\gab ) \gam \chi
\nonumber \\
&&\qquad\qquad -[\gaa ,\gab ] f_m^2 \nonumber \\
&&\qquad\qquad +\frac 12 f_c f_m (-2\gaa \gac \gab
+2\gab\gac\gaa +\gac [\gaa , \gab ] +[\gaa ,\gab ]\gac )
\nonumber \\
&&E=\chi f_{a,a} -(m-1)f_m^2 +2(m-3) \gaa\gam f_a f_m.
\end{eqnarray*}
Since we are looking for the terms with
$f_m^2 f_{a,a}$, all derivatives with respect to the 
$m$-th coordinate are not considered. We compute:
\begin{eqnarray*}
&&\chi_{:a}=-2f_m \gaa \gam \nonumber \\
&&\chi_{:ab}=-2\gaa \gam f_{m,b} 
+([\gac ,\gab] \gaa +\gaa [\gac ,\gab ] )\gam \chi f_c f_m
+4\delta_{ab} \chi f_m^2 \nonumber \\
&&\chi_{:aa}=-2\gaa \gam f_{m,a} +4(m-1) f_m^2 \chi.
\end{eqnarray*}
Only four invariants contain $f_m^2f_{a,a}$:

\medskip

\centerline{\begin{tabular}{|l|l|l|}
\hline 
Invariant & Coefficient of $f_m^2 f_{a,a}$ & Coefficient in $a_5$ \\
\hline 
$\chi \chi_{:a}\chi_{:b} \Omega_{ab}$ & $8(m-2)$  & $w_{14}$ \\
$\chi E^2$                            & $-2(m-1)$ & $720$ \\
$\chi_{:a} E_{:a}  $                  & $-2(m+1)$ & $-180$ \\
$\chi\chi_{:a}\chi_{:a} E$            & $-4(m-1)$ & $-90$ \\
\hline
\end{tabular}}
\medbreak Since all coefficients before odd powers of $f_i$ must vanish, 
$w_{14}=90$. This completes the proof of Lemma \ref{Sectga} and thereby of the
main result of this paper, Theorem \ref{ThmB}.
\qedbox
\appendix
\section{Appendix}

We adopt the notation of \S\ref{Sectd}. Let\smallbreak\noindent
\centerline{\begin{tabular}{ll}
$  M=S^{ 1} \times [0,1]$,&$  \chi :=\sqrt{-1}i$,\\
$ A:=a_{ 0}+ia_1+ja_{ 2}+ka_{3}$,&
$ B:=b_{ 0}+ib_1+jb_{ 2}+kb_{3}$,\\
$ P:=i \partial_1+j \partial_{2}+A+B$,&
$ P^{ *}:=i \partial_1+j \partial_{2}+A-B$,\\
$ D=P^{ *}P$,&\end{tabular}}
A section $f$ satisfies the boundary conditions ($\cab f=0$) if and only if we
have
$\Pi_{-}f |_{ 
\partial M}=0$ and
$  \Pi_{ -}Pf |_{  \partial M}=0$.

\begin{lemma}\label{A1} We have the following relations:
\par
$  \omega_1=a_1-a_{ 0}i+b_{
3}j-b_{ 2}k$ and
$  \omega_{ 2}=a_{ 2}-a_{ 0}j-b_{
3}i+b_1k$.
\par
$  \Omega_{ 12}=(\dot a_{ 2}-  \tilde a_1)+(-\dot b_{ 3}+  \tilde a_{ 0}+2b_{
3}b_1-2a_{ 0}b_{ 2})i+(-\dot a_{
0}-  \tilde b_{ 3}+2a_{ 0}b_1+2b_{
2}b_{ 3})j $\smallbreak\qquad\qquad
$ +(\dot b_1+  \tilde b_{ 2}+2a_{
0}^2+2b_{ 3}^2)k $.
\par
$ E=(\dot b_1+  \tilde b_{ 2}+a_{
0}^2+a_{ 3}^2+b_{ 0}^2+b_{ 3}^2)+(-\dot b_{ 0}-  \tilde a_{
3}+2a_{ 3}b_{ 2}+2b_{ 0}b_1)i $\smallbreak\qquad\qquad
$ +(\dot a_{ 3}-  \tilde b_{ 0}-2a_{
3}b_1+2b_{ 0}b_{ 2})j+(-\dot a_{
2}+  \tilde a_1-2a_{ 0}a_{ 3}+2b_{
0}b_{ 3})k$.
\par
$  \chi_{ :1}=\sqrt{-1}(-2b_{ 3}k-2b_{2}j)$,
$  \chi  \chi_{ :1}=2(b_{ 2}k-b_{3}j) $ and
$  \chi_{ :1}^2=4b_{ 3}^2+4b_{ 2}^2$.
\par
$  \chi_{ :11}=\sqrt{-1}((-4b_{ 3}^2-4b_{ 2}^2)i+(-2\dot b_{ 2}-4a_{
0}b_{ 3})j+(-2\dot b_{ 3}+4a_{
0}b_{ 2})k)$.
\par
$ S= \Pi_{ +}(\sqrt{-1}a_{ 3}+b_{
2})$.
\par
$ S_{ :1}= \Pi_{ +}(\sqrt{-1}\dot a_{
3}+\dot b_{ 2})+\sqrt{-1}(\sqrt{-1}a_{
3}+b_{ 2})(-b_{ 3}k-b_{ 2}j)$.
\par
$  \chi_{ :1}S-S \chi_{ :1}=2(\sqrt{-1}a_{
3}+b_{ 2})(b_{ 3}j-b_{ 2}k) $.
\end{lemma}

\medbreak\noindent{\bf Proof:}  We compute that:

\par\noindent
\PGa
$ D=P^{ *}P=- \partial_1^2- \partial_{ 2}^2+(iA+Ai+iB-Bi) \partial_1+(jA+Aj+jB-Bj) \partial_{ 2 }$

\par\noindent
\PGc
$ +i\dot A+i\dot B+j  \tilde A+j  \tilde B+AB-BA+A^2-B^{ 2 }$

\par\noindent
\PGa
$  \omega_1=- {\textstyle{1\over2}} (iA+Ai+iB-Bi)=a_1-a_{ 0}i+b_{ 3}j-b_{ 2}k $

\par\noindent
\PGa
$  \omega_{ 2}=- {\textstyle{1\over2}} (jA+Aj+jB-Bj)=a_{
2}-a_{ 0}j-b_{ 3}i+b_1k $

\par\noindent
\PGa
$  \omega_1 \omega_{ 2}- \omega_{
2} \omega_1=2a_{ 0}^2k+2a_{
0}b_1j+2b_{ 3}^2k+2b_{
3}b_1i-2a_{ 0}b_{ 2}i+2b_{
2}b_{ 3}j $

\par\noindent
\PGa
$  \Omega_{ 12}= \partial_1 \omega_{
2}- \partial_{ 2} \omega_1+ \omega_1 \omega_{ 2}- \omega_{ 2} \omega_{
1 }$

\par\noindent
\PGb
$ =(\dot a_{ 2}-  \tilde a_1)+(-\dot b_{
3}+  \tilde a_{ 0}+2b_{ 3}b_1-2a_{
0}b_{ 2})i+(-\dot a_{ 0}-  \tilde b_{
3}+2a_{ 0}b_1+2b_{ 2}b_{
3})j $

\par\noindent
\PGc
$ +(\dot b_1+  \tilde b_{ 2}+2a_{
0}^2+2b_{ 3}^2)k $

\par\noindent
\PGa
$ E= {\cal E}_1+ {\cal E}_{ 2}$
for

\par\noindent
\PGa
$  {\cal E}_1=-i\dot A-i\dot B-j  \tilde A-j  \tilde B-\dot  \omega_1-  \tilde  \omega_{ 2 }$

\par\noindent
\PGb
$ =-i\dot A-i\dot B+ {\textstyle{1\over2}} (i\dot A+\dot Ai+i\dot B-\dot
Bi)-j  \tilde A-j  \tilde B+ {\textstyle{1\over2}} (j  \tilde A+  \tilde
Aj+j  \tilde B-  \tilde Bj) $

\par\noindent
\PGb
$ = {\textstyle{1\over2}} (\dot Ai-i\dot A-i\dot B-\dot Bi)+ {\textstyle{1\over2}}
(  \tilde Aj-j  \tilde A-j  \tilde B-  \tilde Bj) $

\par\noindent
\PGb
$ =\dot a_{ 3}j-\dot a_{ 2}k+\dot b_1-\dot b_{ 0}i-  \tilde a_{ 3}i+  \tilde a_1k+  \tilde b_{ 2}-  \tilde b_{ 0}j $

\par\noindent
\PGa
$  {\cal E}_{ 2}=B^2-A^2+BA-AB- \omega_1^2- \omega_{ 2}^{ 2 }$

\par\noindent
\PGa
$ =b_{ 0}^2-b_1^2-b_{
2}^2-b_{ 3}^2+2b_{ 0}b_1i+2b_{ 0}b_{ 2}j+2b_{ 0}b_{
3}k $

\par\noindent
\PGc
$ -a_{ 0}^2+a_1^2+a_{
2}^2+a_{ 3}^2-2a_{ 0}a_1i-2a_{ 0}a_{ 2}j-2a_{ 0}a_{
3}k $

\par\noindent
\PGc
$ +2(a_{ 2}b_1-a_1b_{
2})k+2(a_{ 3}b_{ 2}-a_{ 2}b_{
3})i+2(a_1b_{ 3}-a_{ 3}b_1)j $

\par\noindent
\PGc
$ -a_1^2+a_{ 0}^2+b_{
3}^2+b_{ 2}^2+2a_1a_{
0}i-2a_1b_{ 3}j+2a_1b_{
2}k $

\par\noindent
\PGc
$ -a_{ 2}^2+a_{ 0}^2+b_{
3}^2+b_1^2+2a_{ 2}a_{
0}j+2a_{ 2}b_{ 3}i-2a_{ 2}b_1k $

\par\noindent
\PGa
$ =a_{ 0}^2+a_{ 3}^2+b_{
0}^2+b_{ 3}^2-2a_{ 0}a_{
3}k+2a_{ 3}b_{ 2}i-2a_{ 3}b_1j $

\par\noindent
\PGc
$ +2b_{ 0}b_1i+2b_{ 0}b_{
2}j+2b_{ 0}b_{ 3}k $

\par\noindent
\PGa
$  \chi_{ :1}=\sqrt{-1}[ \omega_1,i)]=\sqrt{-1}[-a_{ 0}i+b_{ 3}j-b_{
2}k,i] $

\par\noindent
\PGb
$ =2\sqrt{-1}(-b_{ 3}k-b_{ 2}j) $

\par\noindent
\PGa
$  \chi_{ :11}= \partial_1 \chi_{
:1}+[ \omega_1, \chi_{ :1}] $

\par\noindent
\PGb
$ =\sqrt{-1}[-2\dot b_{ 3}k-2\dot b_{
2}j-4a_{ 0}b_{ 3}j+4a_{ 0}b_{
2}k-4b_{ 3}^2i-4b_{ 2}^2i]$. 

\par\noindent
Suppose
$  \Pi_{ -}f=0$ on the boundary. Then

\par\noindent
\PGa
$  \Pi_{ -}Pf= \Pi_{ -}(i \partial_1+j \partial_{ 2}-j(jA+jB))f $

\par\noindent
\PGb
$ =j \Pi_{ +}( \partial_{ 2}+ \omega_{
2}- \omega_{ 2}-jA-jB) \Pi_{ +}f $

\par\noindent
\PGa
$ S= \Pi_{ +}( {\textstyle{1\over2}} (jA+Aj+jB-Bj)-jA-jB) \Pi_{
+ }$

\par\noindent
\PGb
$ = {\textstyle{1\over2}}  \Pi_{ +}(-jA+Aj-jB-Bj) \Pi_{
+}$
\par\noindent
\PGb$= \Pi_{ +}(-a_{ 3}i+a_1k+b_{
2}-b_{ 0}j) \Pi_{ + }$

\par\noindent
\PGb
$ = \Pi_{ +}(-ia_{ 3}+b_{ 2})= \Pi_{
+}(\sqrt{-1}a_{ 3}+b_{ 2}) $

\par\noindent
\PGa
$ S_{ :1}= \Pi_{ +}(\sqrt{-1}\dot a_{
3}+\dot b_{ 2})+ {\textstyle{1\over2}}  \chi_{
:1}(\sqrt{-1}a_{ 3}+b_{ 2})$.
\smallbreak\noindent The Lemma now follows. \qedbox

\medbreak
If
$  {\cal E} $ is a local invariant, let
$  \mu ( {\cal E} )=\ptr( {\cal E} (D, \cab ))-\ptr( {\cal E} (  \hat
D,  \hat B))$. Then $
\mu (a_n(1, \cdot ))[\partial M]=0$. 
We compute
$  \mu ( {\cal E} )$ for the terms appearing
in the formula for $a_5$ contained in \S\ref{Sectd}. We organize these
terms into two lemmas to group the data involved.
In Lemma \ref{A2}, we determine the terms which
are bilinear in
$     A$ and
$     B$ or which involve
$   \tilde a_{ 0}a_{ 0}b_{ 2}$.
In Lemma \ref{A3}, we study terms in
$ a_{ 3}$ and in the jets of
$     B$.

\begin{lemma}\label{A2} Terms which are bilinear
in
$      A$ and
$     B$, and the term
$  a_{ 0}b_{ 2}  \tilde a_{ 0} {\cal J}
 $.
\par
$   \mu ( \Omega_{ 12}^2)=2(  \tilde a_{
0}\dot b_{ 3}-\dot a_{ 0}  \tilde b_{
3}) {\cal J} +4a_{ 0}b_{ 2}  \tilde a_{
0} {\cal J} +\dots$ \hfill
$ (w_8) $
\par
$  \mu ( \chi  \Omega_{ 12}^2)=2 \{ \dot a_{
2}\dot b_{ 3}-  \tilde a_1\dot b_{
3} \}  {\cal K} +\dots$ \hfill
$ (w_9) $
\par
$  \mu ( \chi  \Omega_{ 12} \chi  \Omega_{
12})=(2  \tilde a_{ 0}\dot b_{ 3}+2\dot a_{
0}  \tilde b_{ 3}+4a_{ 0}b_{ 2}  \tilde a_{
0}) {\cal J} +\dots$ \hfill
$ (w_{10}) $
\par
$  \mu ( \chi E^2)=(2  \tilde a_{
3}\dot b_1+2  \tilde a_{ 3}  \tilde b_{
2}) {\cal K} +\dots$ \hfill $(720)$
\par
$  \mu ( \chi E \chi E)=-2(  \tilde a_{
3}\dot b_{ 0}+\dot a_{ 3}  \tilde b_{
0}) {\cal J} +\dots$ \hfill
$ ( w_2) $
\par
$  \mu (E^2)=2(-  \tilde a_{ 3}\dot b_{
0}+\dot a_{ 3}  \tilde b_{ 0}) {\cal J} +\dots$
\hfill
$ ( w_1) $
\par
$  \mu ( \chi_{ :11}E)=2(\dot a_{
3}\dot b_{ 2}-\dot a_{ 2}\dot b_{
3}+  \tilde a_1\dot b_{ 3}) {\cal K} +\dots$
\hfill
$ ( 240 - w_{18}+w_{21}) $
\par
$  \mu ( \chi_{ :11} \chi_{ :11})=0+\dots$
\hfill
$ ( w_{34}+w_{35}-w_{37}) $
\par
$  \mu ( \chi E_{ :22})=2(- \partial_{
2}^2(a_{ 3}b_{ 2})+2b_1 \partial_{
2} \partial_1a_{ 3}+\dot a_{
3}  \tilde b_1) {\cal K} +\dots$. \hfill
$(360)$
\par
$  \mu  \{ S_{ :1}S_{ :1} \} =\dot a_{
3}\dot b_{ 2} {\cal K} +\dots$. \hfill $(-360)$
\par
$  \mu (SE_{ :2})= {\textstyle{1\over2}} (a_{
3} \partial_1 \partial_{ 2}b_1+a_{ 3} \partial^2_{ 2}b_{
2}+b_{ 2} \partial^2a_{ 3}) {\cal K} \hfill$
$(1440)$
\par\noindent
\qquad\qquad
$ (- {\textstyle{1\over2}} a_{ 3} \partial_1 \partial_{ 2}b_{ 0}+  \tilde a_{
0}a_{ 0}b_{ 2}) {\cal J} +\dots$
\par
$  \mu (S^2E)=0+\dots$ \hfill $(2880)$
\par
$  \mu (S^{ 4})=0+\dots$ \hfill $(1440)$
\par
$  \mu ( \Pi_{ -}S_{ :1}S_{ :1})=0+\dots$
\hfill
$ ( -2w_4) $
\par
$  \mu ( \chi_{ :1} \chi_{ :1}E)=0+\dots$
\hfill
$ ( w_{19}) $
\par
$  \mu ( \chi  \chi_{ :1} \chi_{
:1}E)=0+\dots$ \hfill
$ ( w_{20}) $
\par
$  \mu ( \chi_{ :1}^{ 4})=0+\dots$ \hfill
$ ( w_{32} +w_{33} ) $
\par
$  \mu ((S \chi_{ :1}- \chi_{ :1}S) \Omega_{
12})=0+\dots$ \hfill
$ ( w_{11}) $
\par
$  \mu (( \chi  \chi_{ :1} \Omega_{
12:2}))=-2b_{ 3} \partial_{ 2} \partial_1a_{ 0} {\cal J} -12a_{ 0}b_{
2}  \tilde a_{ 0} {\cal J} +\dots$ \hfill
$ ( w_{15}) $
\end{lemma}

\begin{lemma}\label{A3}
$  a_{ 0}=a_1=a_{ 2}=0$, let
$ a_{ 3}$ be constant, and let
$ B=B(x_1)$.
\par
$   \mu ( \Omega_{ 12}^2)=(4\dot b_{
3}b_{ 3}b_1-4\dot b_1b_{
3}^2) {\cal J} $ \hfill
$ ( w_8) $
\par
$  \mu ( \chi  \Omega_{ 12}^2)=0$.
\hfill
$ ( w_9=180) $
\par
$  \mu (( \chi  \Omega_{ 12})^2)=(4\dot b_{
3}b_{ 3}b_1+4\dot b_1b_{
3}^2) {\cal J} $ \hfill
$ ( w_{10}) $
\par
$  \mu (( \chi E)^2)= \{ 2\dot b_1(a_{ 3}^2+b_{ 0}^2+b_{
3}^2)-4(-\dot b_{ 0}+2a_{ 3}b_{
2})b_{ 0}b_1 $ \hfill $ $ 
\par 
$\phantom{\mu (( \chi E)^2)=}
-8a_{ 3}b_1b_{
0}b_{ 2} \}  {\cal J} $ \hfill
$ ( w_2) $
\par
$  \mu (E^2)= \{ 2\dot b_1(a_{
3}^2+b_{ 0}^2+b_{ 3}^2)+4\dot b_{ 0}b_{
0}b_1 \}  {\cal J} $ \hfill
$ ( w_1) $
\par
$  \mu ( \chi E^2)= \{ -2(a_{ 3}^2+b_{ 0}^2+b_{ 3}^2)(-\dot b_{
0}+2a_{ 3}b_{ 2})$ \hfill  $ $
\par
$\phantom{ \mu ( \chi E^2)=}
-4\dot b_1b_{
0}b_1 \}  {\cal K} $ \hfill $(720)$
\par
$  \mu ( \chi_{ :11}E)= \{ 4(-\dot b_{
0}+2a_{ 3}b_{ 2})(b_{ 3}^2+b_{
2}^2)+4b_{ 0}b_{ 2}\dot b_{
2}$ \hfill $ $
\par
$\phantom{ \mu ( \chi_{ :11}E)=}
+4b_{ 0}b_{ 3}\dot b_{ 3} \}  {\cal K} $
\hfill
$ ( 240 -w_{18} +w_{21}) $
\par
$  \mu ( \chi_{ :11} \chi_{ :11})=0$
\hfill
$ ( w_{34} + w_{35} -w_{37}) $
\par
$  \mu ( \chi E_{ :22})=(-4b^2_1\dot b_{ 0}+8b^2_1a_{
3}b_{ 2}) {\cal K} $ \hfill $(360)$
\par
$  \mu  \{ S_{ :1}S_{ :1} \} = \{ 2a_{
3}b_{ 2}(b_{ 3}^2+b_{ 2}^2) \}  {\cal K} $ \hfill $(-360)$
\par
$  \mu (SE_{ :2})= \{ -2a_{ 3}b_1^2b_{ 2} \} {\cal K} -\{ 2a_{ 3}b_{
0}b_1b_{ 2} \}  {\cal J} $ \hfill
$(1440)$
\par
$  \mu (S^2E)= \{  {\textstyle{1\over2}} (b_{
2}^2-a_{ 3}^2)\dot b_1+2a_{ 3}b_{ 0}b_1b_{ 2} \}  {\cal J} $
\hfill $(2880)$
\par\noindent
\qquad\qquad
$  \{ - {\textstyle{1\over2}} (b_{ 2}^2-a_{ 3}^2)(-\dot b_{ 0}+2a_{
3}b_{ 2})+a_{ 3}b_{ 2}(a_{ 3}^2+b_{ 0}^2+b_{ 3}^2) \}  {\cal K}  $
\par
$  \mu (S^{ 4})=(2a_{ 3}b_{ 2}^{
3}-2a_{ 3}^{ 3}b_{ 2}) {\cal K} $
\hfill $(1440)$
\par
$  \mu ( \Pi_{ -}S_{ :1}S_{ :1})=a_{
3}b_{ 2}(b_{ 3}^2+b_{ 2}^2) {\cal K} $ \hfill
$ ( -2 w_4) $
\par
$  \mu ( \chi_{ :1} \chi_{ :1}E)=(4b_{
3}^2+4b_{ 2}^2)\dot b_1 {\cal J} $ \hfill
$ ( w_{19}) $
\par
$  \mu ( \chi  \chi_{ :1} \chi_{
:1}E)=-4(b_{ 3}^2+b_{ 2}^2)(-\dot b_{ 0}+2a_{ 3}b_{ 2}) {\cal K} $
\hfill
$ ( w_{20})$
\par
$  \mu ( \chi_{ :1}^{ 4})=0$. \hfill
$ ( w_{32} + w_{33})$
\par
$  \mu ((S \chi_{ :1}- \chi_{ :1}S) \Omega_{
12})= -2b_{
2}^2\dot b_1 {\cal J} $
\hfill($ w_{11}) $
\par
$  \mu ( \chi  \chi_{ :1} \Omega_{
12:2})=-4\dot b_{ 3}b_{ 3}b_1+4\dot b_1b_{ 3}^2$. \hfill
$ ( w_{15}) $\end{lemma}
\section{Appendix}
Let $ds^2(\epsilon)=e^{2\epsilon F}ds^2$ define a conformal variation.
In this appendix, we summarize the computation of the conformal variations
we needed in Section \ref{Sectf}. We integrate by parts where necessary to
bring the variations into standard form so that $F$ is not differentiated with
respect to tangential coordinates. We will be dealing with terms which are
homogeneous of order $4$. If $F$ is constant, then $\frac d {d
\epsilon} |_{\epsilon =0}X=-4FX$.  To avoid writing the
conformal weight term repeatedly, we
define ${\cal L}X:=\frac d {d
\epsilon}X |_{\epsilon =0}+4 FX$. Note that ${\cal L}X=0$ if $X$ only involves
$\chi$ and $\Omega$ terms; consequently these terms have
been omitted in the interests of brevity. \smallbreak\noindent
\begin{tabular}{lll}
           & \underline{\mbox{Term}} &$ {\cal L}$\\ 
$w_1 $&$ E^2 $\hglue 1.8cm\ &$ (m-2) \big(
            FE_{:aa} -\fm EL_{aa}+\fmm E \big) $\\
$w_2 $&$ \chi E \chi E $&$ (m-2) \big(
           F E_{:aa} -\fm EL_{aa}+\fmm E\big)  $\\
$w_3 $&$ S_{:a} S_{:a}$&$2F S_{:a}  S_{:a} +2F
       S S_{:aa} -(m-2) \fm S_{:aa}$\\&&$  \quad+\frac {m-2}2\fm \ca S_{:a} $\\
$w_4 $&$ \chi S_{:a}S_{:a}$&$2F  S_{:a}S_{:a}+2FS  S_{:aa} 
  -(m-2) 
    \fm  S_{:aa} $\\
$w_{11} $&$ \Om_{am} [\chi,S_{:a}]$&$ \frac{m-2}2 F_{;m} 
               \chi \chi_{:a} \Om_{am} $\\
$w_{12} $&$ \chi \chi_{:a} \Om_{am} L_{cc} $&$ 
        -(m-1)F_{;m} \chi \chi_{:a} \Om_{am} $\\ 
$w_{15} $&$ \chi \chi_{:a} \Om_{am;m} $&$ 
     F\chi_{:a} \chi_{:b} \Om_{ab} -
                   F\chi \chi_{:a} \Om_{ab:b} -\frac 1 2 F 
      \Om_{ab} \Om_{ab} $\\
        &&$ +\frac 1 2 F\chi \Om_{ab} \chi \Om_{ab} -2\chi 
               \chi_{:a} \Om_{am} F_{;m} $\\
$w_{16} $&$ \chi \chi_{:a} \Om_{ab:b} $&$ 
     (m-5) F \chi_{:a} \chi_{:b}
           \Om_{ab} -(m-5) \chi \chi_{:a} \Om_{ab:b} $\\ 
    &&$+      \frac{m-5}2F(- \Om_{ab} \Om_{ab} +  \chi 
       \Om_{ab} \chi \Om_{ab}) $\\
$w_{17} $&$ \chi \chi_{:a} \Om_{bm} L_{ab} $&$ 
        -\chi \chi_{:a} \Om_{am}F_{;m} $\\
$w_{18} $&$ \ca E_{:a} $&$2F \chi_{:aa}E+2F\ca 
                    E_{:a} $\\
$w_{19} $&$ \ca \chi_{:a} E $&$ \frac{m-2}2 \ca \chi_{:a} \fmm
            -\frac{m-2}2 \ca \chi_{:a} L_{cc} \fm $\\
 &&$ +(m-2) \chi_{:ab} \chi_{:ab} F
         +(m-2) \chi_{:bba} 
                 \chi_{:a} F$\\&&$ +4(m-2) \ca \chi_{:b} \Om_{ab}F
       -2(m-2) \chi \ca \Om_{ab:b} F$\\&&$\  -
           (m-2) \chi_{:a} \chi_{:b} F ( -\rho_{ab} +R_{mbam}
     - L_{ab} L_{cc} +L_{ac} L_{bc}) $\\
$w_{20} $&$ \chi \ca\ca E$ &  $0 $ \\
$w_{21} $&$ \chi_{:aa} E $\hglue 1.1cm\ 
      &$
      -(m-3) F \chi_{:aa}E -(m-3) F \ca E_{:a} $\\
$w_{22} $&$ \ca \chi_{:a} \tau $&$ 
        -2(m-1) \ca \chi_{:a} \fmm+2(m-1) \ca \chi_{:a} L_{cc} \fm $\\
      &&$ -4(m-1) \chi_{:ab} \chi_{:ab} F 
     -4 (m-1) \chi_{:bba} \chi_{:a} F$\\&&$ -16 (m-1) 
               \ca \chi_{:b} \Om_{ab} F 
    +8(m-1) \chi \chi_{:a} \Om_{ab:b} F$\\&&$ +
     4(m-1) \chi_{:a} \chi_{:b} F \big(R_{mbam} -\rho_{ab}  -
              L_{ab} L_{cc} +L_{ac} L_{bc}\big)$\\
$w_{23} $&$ \ca \chi_{:a} \rho_{mm} $&$  -(m-1) \ca \chi_{:a} 
    \fmm +\ca \chi_{:a} L_{cc} \fm -2\chi_{:ab} \chi_{:ab} F $\\
  &&$         -2 \chi_{:bba} \chi_{:a} F -8 \ca \cb \Om_{ab} F 
        +4\chi \ca \Om_{ab:b} F $\\
  &&$ +2F\chi_{:a} \chi_{:b} 
              \big( -\rho_{ab} +R_{mbam} -
              L_{ab} L_{cc} +L_{ac} L_{bc}\big) $
\end{tabular}\smallbreak\noindent
\begin{tabular}{lll}
$w_{24} $&$ \ca \ca \rho_{ab} $&$  -\ca \chi_{:a} \fmm 
      +\ca \chi_{:a} L_{cc} \fm +(m-2) \ca \cb L_{ab} \fm $\\
 &&$ -m \chi_{:ab} \chi_{:ab} F
      +2(1-m) \chi_{:aab} \chi_{:b} F$\\&&$ 
      -(2m+4) F \ca \cb \Om_{ab}     
      -(m-2) \chi_{:aa} \chi_{:bb} F$\\&&$ +4F \chi \ca 
     \Om_{ab:b} -(m-2) F \Om_{ab} \Om_{ab} $\\ 
    &&$   +(m-2)F \chi \Om_{ab} \chi \Om_{ab}$\\&&$ +m F \chi_{:a} \chi_{:b} 
          \big[ -\rho_{ab} +R_{mbam} -
              L_{ab} L_{cc} +L_{ac} L_{bc}\big] $\\
$w_{25} $&$ \ca \cb R_{mabm} $&$ 
     -\ca \chi_{:a} \fmm +\ca \cb 
     L_{ab} \fm -2F\chi_{:aab} \chi_{:b}$\\
  &&$ - 
    2F\ca \cb \Om_{ab} -F \chi_{:ab} \chi_{:ab} -
    F\chi_{:aa} \chi_{:bb} -F\Om_{ab}\Om_{ab}$\\&&$ +
    F\chi \Om_{ab} \chi \Om_{ab} 
    +F\chi_{:a} \chi_{:b} 
           \big[ -\rho_{ab} +R_{mbam} -
              L_{ab} L_{cc}$\\&&$ +L_{ac} L_{bc}\big] $\end{tabular}\smallbreak\noindent
\begin{tabular}{lll}
$w_{26} $&$ \ca \chi_{:a} L_{bb}L_{cc} $&$ -2(m-1) \fm L_{cc} \ca \chi_{:a} $\\
$w_{27} $&$ \ca \cb L_{ac} L_{bc} $&$  -2\ca \cb 
    L_{ab} \fm $\\
$w_{28} $&$ \ca \chi_{:a} L_{cd} L_{cd} $&$  -
         2\fm L_{cc} \ca \chi_{:a} $\\ 
$w_{29} $&$ \ca \cb L_{ab} L_{cc} $&$  -\ca \chi_{:a} L_{cc} \fm 
         -(m-1) \ca \cb L_{ab} \fm $\\
$w_{30} $&$ \ca S_{:a}L_{bb}$&$\frac 1 4 (m-2) \fm \ca \chi_{:a} L_{cc} 
    -(m-1) \fm \ca S_{:a}$\end{tabular}\smallbreak\noindent
\begin{tabular}{lll}
$w_{31} $&$ \ca S_{:b} L_{ab} $\hglue .6cm\ &$  -\ca S_{:a}\fm 
    +\frac 1 4 (m-2) \fm \ca \cb L_{ab} $\\
$w_{34} $&$ \chi_{:aa} \chi_{:bb} $&$-
    2(m-3) F\chi_{:aa} \chi_{:bb} -2(m-3) 
           F\ca \chi_{:bba} $\\
$w_{35} $&$ \chi_{:ab} \chi_{:ab} $&$
            4 F \chi_{:ab} \chi_{:ab} +2F \Om_{ab} \Om_{ab} -
         2F\chi \Om_{ab} \chi \Om_{ab}$\\&&$ -2F \chi_{:aa} 
            \chi_{:bb} 
        +2F \chi_{:b} \chi_{:aab} 
            -4F\chi_{:a} \chi_{:b}  \big[ -\rho_{ab} +R_{mbam}$\\&&$ -
              L_{ab} L_{cc} +L_{ac} L_{bc}\big] 
              +12 F \ca \cb \Om_{ab} 
        -4F \chi \chi_{:a} \Om_{ab:b} $\\
$w_{36} $&$ \ca \chi_{:a} \chi_{:bb} $&$ 0$\\
$w_{37} $&$ \cb \chi_{:aab} $&$
           (m-1) F \cb \chi_{:aab} +
         (m-1) F \chi_{:bb} \chi_{:aa} 
$\end{tabular}
\section{Appendix}
In the proof of Lemma \ref{Sectga}, we used the local
index theorem.
We list in this appendix  the terms giving non-zero contributions
to $a_5(D^{[1]}) - a_5(D^{[0]})$. Coefficients
of corresponding invariants are given an the right hand side
in the formulae below. Let a comma denote ordinary partial differentiation. The terms
with
$f_{,m}f_{,mmm}$ are dropped since they are not needed for the calculations.
We omit the factor of $\ptr_V I_V$.
\smallbreak\noindent\begin{tabular}{rll}
\underline{\mbox{Coef}}&\underline{\mbox{Term}}\\
1440&$E_{;m}S$ &$ \frac 12 (m-1)^2 (\frac 12 (m^2 -8) f_{,m}^2f_{,mm}$\\
          &&$-f_{,am}f_{,am}-\frac 12 (m^2-4)f_{,m}^4$\\
480&$\tau S^2 $ & $\frac 12 (m-1)^3 f_{,m}^2 (-2f_{,mm}-(m-2)f_{,m}^2)$ \\
270&$\tau_{;m}S$ & $(m-1)^2 (f_{,am}f_{,am} -(m-4)f_{,mm}f_{,m}^2+(m-2)f_{,m}^4 )$\\
120&$\rho_{mm}S^2$ & $-\frac 12 (m-1)^3 f_{,m}^2f_{,mm}$  \\
1080&$SS_{:aa}$ & $-\frac 1 2 (m-1)^2 
   (\aam \aam +\frac 1 2 m^2(m-1) \am^4 )$\\
2880&$ES^2$ & $\frac 12 (m-1)^3 (f_{,m}^2 f_{,mm} +\frac 14 (m^2-4) f_{,m}^4)$\\
1440&$S^4$ & $\frac 12 (m-1)^4 f_{,m}^4$ 
\end{tabular}\smallbreak\noindent\begin{tabular}{rll}
\underline{\mbox{Coef}}&\underline{\mbox{Term}}\\
270&$L_{aa}E_{;m}$ & $-(m-1)^2 (2(m-3)f_{,m}^2f_{,mm}-2(m-2)f_{,m}^4$\\
       &&$-f_{,am}f_{,am})$\\ 
270&$L_{bb}S_{:aa}$ & $\frac 1 2 (m-1)^2
      f_{,ma}f_{,ma}$\\ 
1440&$L_{bb}SE$ & $-\frac 12 (m-1)^3 (f_{,m}^2f_{,mm} +\frac 14
     (m^2-4)f_{,m}^4)$\\ 
30&$L_{bb}S\rho_{mm}$ & $\frac 12 (m-1)^3 f_{,m}^2 f_{,mm}$\\
240&$L_{bb}S\tau $ & $\frac 12 (m-1)^3 (2f_{,m}^2f_{,mm} +(m-2)f_{,m}^4)$ \\
-60&$L_{ab}\rho_{ab}S$ & $-\frac 12 (m-1)^2 (-f_{,m}^2f_{,mm}
                       +(2-m)f_{,m}^4)$ \\
180&$L_{ab}SR_{ammb}$ & $\frac 12 (m-1)^2 f_{,m}^2f_{,mm}$\\
45&$L_{aa}L_{bb} E$ & $(m-1)^3 (\am^2 \amm +(m-2) \am^4 )$\\
90&$L_{ab} L_{ab} E $ & $(m-1)^2 (\am^2 \amm +(m-2) \am^4 )$\\
2160&$L_{bb}S^3$ & $-\frac 1 2 (m-1)^4 \am^4$\\
1080&$L_{aa}L_{bb}S^2$ & $\frac 1 2 (m-1)^4 \am^4$\\
360&$L_{ab} L_{ab} S^2$ & $\frac 1 2 (m-1)^3 \am^4$\\
$\frac{885} 4 $&$L_{aa}L_{bb}L_{cc} S$ & $-\frac 1 2 (m-1)^4 \am^4$\\
$\frac{315} 2 $&$L_{cc}L_{ab}L_{ab}S$ & $-\frac 1 2 (m-1)^3 \am^4$\\
150&$L_{ab}L_{bc}L_{ca}S$ & $-\frac 1 2 (m-1)^2 \am^4$\\
$w_{12}$ &$\chi \ca \Om_{am} L_{bb}$ & $2(m-1)^3 \am^2\amm$\\
$w_{13}$&$\ca \cb \Om_{ab}$ & $2(m-1)^2(m-2)(m^2+(m-2)^2)\am^4$\\
90&$\chi \ca \Om_{am;m}$ & $2(m-1)(2(m-1)\am^2\amm +\aam\aam)$\\
120&$\chi \ca \Om_{ab:b}$ & $\frac 1 2 (m-2) ((m^2-(m-2)^2) \aam\aam$ \\
     && $+(m-1)(m^4-(m-2)^4) \am^4)$ \\
$w_{17}$&$\chi\ca \Om_{bm} L_{ab}$ & $2(m-1)^2 \am^2 \amm$\\
$-180$&$\ca \ca E$ & $(m-1)^2 (m^2\am^2\amm +\frac 1 4 (m^2(m^2-4)$\\&&$-(m-2)^4) 
    \am^4)$\\
$-30$&$\ca \ca \tau $ & $-4(m-1)^3 (2 \am^2\amm +(m-2)\am^4)$\\
0&$\ca\ca \rho_{mm} $ & $-4(m-1)^3 \am^2\amm$\\
$-60$&$\ca \cb \rho_{ab}$ & $-4(m-1)^2 (\am^2 \amm +(m-2)\am^4)$\\
30&$\ca \cb R_{mabm}$ & $-4(m-1)^2\am^2\amm$\\
 $\frac{255}4$&$\chi_{:aa} \chi_{:bb}$ & $(m-1)(4\aam\aam +(m-1)(m^4-(m-2)^4)\am^4)$\\
$-\frac{105}2$&$\chi_{:ab}\chi_{:ab}$ & $(m-1) (4(m-1)\aam\aam +(m^4-(m-2)^4)\am^4)$\\
$\frac{15} 4$&$\ca \ca\cb\cb$ & $(m-1)^2 (m^4 -(m-2)^4) \am^4$  \\
 $\frac{15} 8$&$\ca \cb\ca\cb$ & $-(m-1) (m-3) (m^4 -(m-2)^4) \am^4$\\
$-\frac{675} 8$&$\ca\cb L_{ab}L_{cc}$ & $4(m-1)^3 \am^4$\\
$-\frac{675}{32}$ &$\ca\ca L_{bb}L_{cc}$ & $4(m-1)^4\am^4$\\
 $-\frac{75} 4$&$\ca \cb L_{ac}L_{cb} $ & $4(m-1)^2 \am^4$ \\
 $-\frac{195}{16}$&$\ca\ca L_{cd}L_{cd}$ & $4(m-1)^3\am^4$ \\
 $-330$ &$\ca S_{:a} L_{bb}$ & $-\frac 1 2 m^2(m-1)^3\am^4$\\
 $-300$&$\ca S_{:b}L_{ab}$ & $-\frac 1 2 m^2(m-1)^2 \am^4$ \\
 720& $\chi S_{:a} S_{:a} $ & $\frac 1 2 (m-1)^2 \aam\aam$
\end{tabular}
\smallbreak\noindent\begin{tabular}{rll}
\underline{\mbox{Coef}}&\underline{\mbox{Term}}\\
0&$\Om_{ab} \Om_{ab}+\chi \Om_{ab}\chi\Om_{ab}$ & $-\frac 1 2 (m-1)(m-2)
        (m^4-(m-2)^4)\am^4$\\
$\frac12(w_5 -w_7)$ &  $\Om_{ab} \Om_{ab}-\chi \Om_{ab} \chi \Om_{ab}$ &
        $-4(m-1)(m-2)\aam \aam$ \\
$-45$&$\Om_{am}\Om_{am}+\chi \Om_{am}\chi \Om_{am} $ & $-2(m-1)(m-2)\aam\aam$ \\
$-360$&$\Om_{am}\chi S_{:a}-\Om_{am}S_{:a}\chi$ & $-\frac 1 2 m^2(m-1)^2 \am^2\amm$
\end{tabular}

\smallbreak\noindent T. Branson, Department of Mathematics, The University of Iowa, 
Iowa City IA 52242 USA. EMAIL: branson@math.uiowa.edu
\smallbreak\noindent P. Gilkey, Department of Mathematics, University of
Oregon, Eugene OR 97403 USA. EMAIL: gilkey@darkwing.uoregon.edu
\smallbreak\noindent K. Kirsten, Department of Physics and Astronomy, The
University of Manchester, Oxford Road, Manchester, England. EMAIL:
klaus@a13.ph.man.ac.uk
\smallbreak\noindent D. Vassilevich, Institute for Theoretical Physics, 
University of Leipzig, Augustusplatz 10, 04109 Leipzig, Germany. EMAIL:
vassil@itp.uni-leipzig.de
\end{document}